\documentclass[seceq,supplement]{ptptex}

\usepackage{graphicx}

\title{
Mott transitions in two-orbital Hubbard systems
}

\author{
Akihisa \textsc{Koga}, 
Kensuke \textsc{Inaba}, 
and 
Norio \textsc{Kawakami}
}

\inst{
Department of Applied Physics, Osaka University,
  Suita, Osaka 565-0871
}

\abst{
 We investigate the Mott transitions in two-orbital Hubbard systems. 
Applying the dynamical mean field theory and the 
self-energy functional approach,
we discuss the stability of itinerant quasi-particle states in each band.
It is shown that separate Mott transitions occur at different
Coulomb interaction strengths in general.
On the other hand, if some special conditions are satisfied for the
interactions, spin and orbital fluctuations are equally enhanced
at low temperatures, resulting in a single Mott transition.
The phase diagrams are obtained at zero and finite temperatures.
We also address the effect of the hybridization between two orbitals, 
which induces the Kondo-like heavy fermion states in the 
intermediate orbital-selective Mott phase.
}

\begin{document}
\ifx\href\undefined\else\hypersetup{linktocpage=true}\fi 

\maketitle

\section{Introduction}
Strongly correlated electron systems with some orbitals
 have been investigated  extensively.\cite{ImadaRev,TokuraScience}
In particular, substantial progress in the theoretical 
understanding of the Mott transition in multiorbital systems 
has been made by dynamical mean-field theory (DMFT)
\cite{Georges,KotliarPT,PruschkeAP,Metzner,Muller} 
calculations. 
\cite{2band1,2band2,Florens,Kotliar96,Rozenberg97,Bunemann:Gutzwiller,Hasegawa98,Held98,Han98,Momoi98,Klejnberg98,Imai01,Koga,KogaSN,Oudovenko02,Ono03,Tomio04,Pruschke,Sakai,Liebsch,KogaLett,KogaB,Ferrero05,Medici05,Arita05,Knecht05,LiebschFiniteT,Biermann05,Inaba,Ruegg,sces,Song,InabaB}
Among them, the orbital-selective Mott transition (OSMT)\cite{Anisimov}
 has been one of the most active topics in this context. 
A typical material is the single-layer isovalent ruthenate alloy
Ca$_{2-x}$Sr$_x$RuO$_4$\cite{Nakatsuji,Nakatsuji1,tilting}. 
The end-member Sr$_2$RuO$_4$ is a well-known
unconventional superconductor \cite{PT,RMP}, while Ca$_2$RuO$_4$ is a
Mott-insulating $S=1$ antiferromagnet \cite{Ca2RuO41,Ca2RuO42,Ca2RuO43}. 
The relevant $4d$-orbitals belong to the $t_{2g}$-subshell.
The planar structure prevents the hybridization between orbitals
which have even ($d_{xy}$) and odd parity ($d_{yz},d_{zx}$) under the
reflection $ z \to -z $. 
The complex evolution between these different end-members has 
stimulated  theoretical investigations \cite{Mazin,Hotta,Fang,Okamoto}, 
and among others to the proposal of the OSMT: some of
the $d$-orbitals fall in localized states, while the
 others provide itinerant electrons. The OSMT scenario
could explain the experimental
observation of a localized spin $S=1/2 $ in the metallic system at $x
\sim 0.5 $ in Ca$_{2-x}$Sr$_x$RuO$_4$, which is difficult 
to obtain from the entirely itinerant description
\cite{Anisimov,Fang,SigristTroyer}.
Another example of the OSMT is the compound 
$\rm La_{n+1}Ni_nO_{3n+1}$.\cite{LaNiO} It is reported that
the OSMT occurs in the $e_g$-subshell
at the critical temperature $T_c\sim 550K$, below which
the conduction of electrons in the $3d_{x^2-y^2}$ orbital is disturbed 
by the Hund coupling with the localized electrons in the $3d_{3z^2-r^2}$
orbital.\cite{Kobayashi96}
These experimental findings have stimulated theoretical 
investigations of the Mott transitions in the multiorbital systems.
\cite{Liebsch,SigristTroyer,Fang,Anisimov,KogaLett,KogaB,Ferrero05,Medici05,Arita05,Knecht05,LiebschFiniteT,Biermann05,sces,Ruegg,Inaba,Tomio04}

In this paper, we give a brief review of our recent studies 
\cite{Koga,KogaSN,KogaLett,KogaB,Inaba,InabaB}
on the Mott transitions in the two-orbital Hubbard model by means of DMFT and 
the self-energy functional approach (SFA).\cite{SFA} 
In particular, we focus on the role of the orbital degrees of freedom 
to discuss the stability of the metallic state
at zero and finite temperatures.
The paper is organized as follows.
In \S\ref{sec2}, we introduce the model Hamiltonian for the two-orbital 
systems and briefly explain the framework of DMFT and SFA. 
We first treat the Hubbard model with same bandwidths, and
 elucidate how enhanced spin and orbital fluctuations affect 
the metal-insulator transition in \S \ref{sec3}.
In \S \ref{sec4}, we then consider the system with different bandwidths 
to discuss in which conditions the OSMT occurs.  We obtain the 
phase diagrams at zero and finite temperatures.
Finally in \S \ref{sec5}, the effect of the hybridization between the 
two orbitals is investigated to clarify the instability of the intermediate 
OSM phase. A brief summary is given in the last section.

\section{Model and Methods}\label{sec2}

\subsection{Two-orbital Hubbard model}

We study the two-orbital Hubbard Hamiltonian,
\begin{eqnarray}
H&=&H_0+H'\\
H_0&=&\sum_{\stackrel{<i,j>}{\alpha\sigma}}
t_{ij}^{(\alpha)} c_{i\alpha\sigma}^\dag c_{j\alpha\sigma}
+V\sum_{i\sigma}\left[c_{i1\sigma}^\dag c_{i2\sigma}+
c_{i2\sigma}^\dag c_{i1\sigma}\right]\\
H'&=&\sum_i H'_i\\
H'_i&=&
U\sum_{\alpha}n_{i\alpha\uparrow}n_{i\alpha\downarrow}
+\sum_{\sigma\sigma'}
\left(U'-J_z\delta_{\sigma\sigma'}\right)n_{i1\sigma}n_{i2\sigma'}
\nonumber\\
&-&J\sum_\sigma c_{i1\sigma}^\dag c_{i1\bar{\sigma}}
c_{i2\bar{\sigma}}^\dag c_{i2\sigma}
-J'\left[ c_{i1\uparrow}^\dag c_{i1\downarrow}^\dag
c_{i2\uparrow} c_{i2\downarrow}+ c_{i2\uparrow}^\dag c_{i2\downarrow}^\dag
c_{i1\uparrow} c_{i1\downarrow} \right]\label{eq:model}
\label{Hamilt}
\end{eqnarray}
where $c_{i\alpha\sigma}^\dag (c_{i\alpha\sigma})$ 
creates (annihilates) an electron 
with  spin $\sigma(=\uparrow, \downarrow)$ and orbital
 $\alpha(=1, 2)$ at the $i$th site and 
$n_{i\alpha\sigma}=c_{i\alpha\sigma}^\dag c_{i\alpha\sigma}$. 
$t_{ij}^{(\alpha)}$ is the orbital-dependent nearest-neighbor hopping, 
$V$ the hybridization between two orbitals and
$U$ ($U'$) represents the intraorbital (interorbital) Coulomb interaction.
$J$, $J_z$ and $J'$ represent the spin-flip, Ising, and pair-hopping 
components of  the Hund coupling, respectively.  We note that
when the system has the Ising type of anisotropy in the Hund 
coupling, $J=J'=0$,
the system at low temperatures should exhibit quite different properties
from the isotropic case.
\cite{sces,Pruschke,Knecht05,Liebsch,Ruegg,LiebschFiniteT,Biermann05,Han98}
In this paper, we deal with the isotropic case and 
set the parameters as $J=J_z=J'$, which satisfy
the symmetry requirement in the multi-orbital systems.
It is instructive to note that this generalized model allows us 
to study a wide variety of different models 
in the same  framework. For $V=0$, 
the system is reduced to the multi-orbital Hubbard model 
with the same $(t_{ij}^{(\alpha)}=t_{ij})$ or distinct orbitals.
\cite{KogaLett,Tomio04,Ruegg,Liebsch,KogaB,sces,Ferrero05,Medici05,Arita05,Knecht05,LiebschFiniteT,Biermann05,Inaba}
On the other hand, for $t_{ij}^{(2)}=0$, the system is reduced to 
a correlated electron system coupled to localized electrons, 
such as the periodic Anderson model ($J=0$),
\cite{Coleman,Rice,Yamada,Kuramoto,Kim90}
the Kondo lattice model ($V=0$ and $J<0$)\cite{Tsunetsugu,Assaad}
for heavy-fermion systems, and
the double exchange model
($V=0$ and $J>0$) for some transition metal oxides. 
\cite{Zener,Anderson,Kubo,Furukawa}
For general choices of the parameters, various
characteristic  properties may show up, which
continuously bridge these limiting cases.

\subsection{Method}

\subsubsection{dynamical mean-field theory}

To investigate the Hubbard model (\ref{Hamilt}),
we make use of DMFT, \cite{Metzner,Muller,Georges,PruschkeAP,KotliarPT}
which has successfully been applied to various electron systems such as 
the single-orbital Hubbard model, 
\cite{Caffarel,2site,LDMFT,single1,Rozenberg1,OSakai,single2,single3,single4,BullaNRG,OnoED,Nishimoto,Uhrig,Zitzler}
the multiorbital Hubbard model, 
\cite{2band1,2band2,Florens,Kotliar96,Rozenberg97,Bunemann:Gutzwiller,Hasegawa98,Held98,Han98,Momoi98,Klejnberg98,Imai01,Oudovenko02,Koga,Ono03,KogaSN,Sakai,Pruschke,Song,InabaB,Liebsch,sces,Knecht05,Biermann05,LiebschFiniteT,Ruegg,KogaLett,KogaB,Tomio04,Ferrero05,Medici05,Arita05,Inaba}
the periodic Anderson model,
\cite{Rozenberg,PAM,Mutou,Saso,Sun,Sato,Ohashi,MediciPAM,SchorkPAM}
the Kondo lattice model,\cite{Matsumoto,OhashiJPC,Schork} etc.
In the framework of DMFT, the lattice model is mapped to
 an effective impurity  model, 
where local electron correlations are taken into account precisely. 
The Green function  for the original lattice system is then obtained 
via self-consistent equations imposed on the impurity problem.

In DMFT for the two-orbital model,
the Green function in the lattice system is given as,
\begin{eqnarray}
{\bf G}\left(k, z\right)^{-1}={\bf G}_0\left(k, z\right)^{-1}
-{\bf \Sigma}\left(z \right),
\end{eqnarray}
with
\begin{equation}
{\bf G}_0\left( k, z\right)^{-1}=\left(
\begin{array}{cc}                 
z+\mu-\epsilon_1( k) & -V\\
-V & z+\mu-\epsilon_2( k)
\end{array}
\right),
\end{equation}
and
\begin{equation}
{\bf \Sigma}\left(z\right)=\left(
\begin{array}{cc}
\Sigma_{11}(z) & \Sigma_{12}(z) \\
\Sigma_{21}(z) & \Sigma_{22}(z) 
\end{array}
\right),
\end{equation}
where $\mu$ is the chemical potential, and $\epsilon_\alpha (k)$
is the bare dispersion relation for the $\alpha$-th orbital. 
In terms of the density of states (DOS) $\rho (x)$ rescaled by the 
bandwidth $D_\alpha$,
the local Green function is expressed as,
\begin{eqnarray}
G_{11}(z)&=&\int dx \frac{\rho(x)}{\xi_1\left(z,x\right)-
\frac{\displaystyle v(z)^2}{\displaystyle \xi_2\left(z, x\right)}},
\nonumber\\
G_{12}(z)&=&\int dx \frac{\rho(x)v(z)}
{\xi_1\left(z, x\right)\xi_2\left(z, x\right)-v(z)^2},
\nonumber\\
G_{22}(z)&=&\int dx \frac{\rho(x)}{\xi_2\left(z, x\right)-
\frac{\displaystyle v(z)^2}{\displaystyle \xi_1\left(z, x\right)}},
\end{eqnarray}
where
\begin{eqnarray}
 \xi_1\left(z, x\right)&=&z+\mu-\Sigma_{11}-D_1 x,\nonumber\\
 \xi_2\left(z, x\right)&=&z+\mu-\Sigma_{22}-D_2 x,\nonumber\\
 v\left(z\right)&=&V+\Sigma_{12}\left(z\right).
\end{eqnarray}
In the following, we use the semicircular DOS,
$\rho(x)=\frac{2}{\pi}\sqrt{1-x^2},$
which corresponds to the infinite-coordination Bethe lattice.

There are various numerical methods 
to solve the effective impurity problem.
We note that self-consistent perturbation theories such as
the non-crossing approximation and the iterative perturbation method
are not efficient enough to discuss  orbital fluctuations
in the vicinity of the critical point.
In this paper, we  use numerical techniques such as
the exact diagonalization (ED) \cite{Caffarel} 
and the quantum Monte Carlo (QMC) simulations\cite{Hirsch}
as an impurity solver at zero and finite temperatures.
In this connection, we note that the Hund coupling is a
 crucial parameter that should  control the nature of 
the Mott transition in the multiorbital systems.\cite{sces}
It is thus important to carefully analyze the effect of 
the Hund coupling in the framework of QMC.
For this purpose, we use the QMC algorithm proposed 
by Sakai et al.,\cite{Sakai} 
where  the Hund coupling is represented in terms of
discrete auxiliary fields.
When we solve the effective impurity model by means of QMC method,
we use the Trotter time slices $\Delta \tau = (TL)^{-1} \le 1/6$,
where $T$ is the temperature and $L$ is the Trotter number.

\subsubsection{self-energy functional approach}

We also make use of a similar but slightly different method, SFA,\cite{SFA}
 to determine the phase diagram at finite temperatures. This SFA, which 
is based on the Luttinger-Ward variational method,\cite{Luttinger}
 allows us to deal with 
finite-temperature properties of the multi-orbital system efficiently,
\cite{Inaba,InabaB} 
where standard DMFT with numerical methods
may encounter some difficulties in practical calculations when 
the number of orbitals increases. 

In SFA, we utilize the fact that the Luttinger-Ward functional 
does not depend on 
the detail of the Hamiltonian ${\cal H}_0$ as far as the interaction term
${\cal H}^\prime$ is unchanged.\cite{SFA}
This enables us to introduce a proper reference system having
the same interaction term.
One of the simplest reference systems is given 
by the following Hamiltonian,
${\cal H}_{\rm ref}=\sum_i{\cal H}_{\rm ref}^{(i)}$, 
\begin{eqnarray}
  {\cal H}_{\rm ref}^{(i)}&=&\sum_{\alpha \sigma } \left[ 
e^{(i)}_{0\alpha}
          c^\dag_{i\alpha\sigma} c_{i\alpha\sigma}+          
e^{(i)}_{\alpha}
          a^{(i)\dag}_{\alpha\sigma}a^{(i)}_{\alpha\sigma}+v^{(i)}_{\alpha}
          \left(c^\dag_{i\alpha\sigma}a^{(i)}_{\alpha\sigma}+a^{(i)\dag}_{\alpha\sigma}c_{i\alpha\sigma}\right)\right]
+H_i^\prime,\label{eq:ref_model}
\end{eqnarray}
where $a^{(i)\dag}_{\alpha\sigma}(a^{(i)}_{\alpha\sigma})$ 
creates (annihilates) an electron with spin $\sigma$ and orbital $\alpha$, 
which is connected to the $i$th site in the original lattice. 
This approximation may be regarded as a finite-temperature extension
of the two-site DMFT\cite{2site}.
In the following, we fix the parameters 
$e_{0\alpha}=0$ and $e_{\alpha}=\mu$
to investigate the zero and finite temperature properties at half filling.
We determine the parameters $v_{\alpha}$ variationally so as 
to minimize the grand potential, 
$\partial\Omega/\partial v_\alpha=0$ $(\alpha=1,2)$, 
which gives a proper reference system within the given form
of the Hamiltonian (\ref{eq:ref_model}).

\section{Mott transition in the degenerate two-orbital model}\label{sec3}

We begin with the two-orbital Hubbard model with same bandwidths 
$(t_{ij}^{(\alpha)}=t_{ij})$ at half filling.
\cite{2band1,2band2,Florens,Kotliar96,Rozenberg97,Bunemann:Gutzwiller,Hasegawa98,Held98,Han98,Momoi98,Klejnberg98,Imai01,Oudovenko02,Koga,KogaSN,Ono03,Sakai,Pruschke,Song,InabaB}
In this and next sections, we neglect the hybridization term by putting $V=0$.
We first discuss the Fermi-liquid properties in the metallic phase 
when the Coulomb interactions $U$ and $U'$ are varied
independently in the absence  of the Hund coupling $(J=0)$.\cite{Koga,KogaSN}
The effect of the Hund coupling will be mentioned 
in the end of this section.\cite{InabaB}

\subsection{zero-temperature properties}

To investigate the Mott transition at zero temperature, 
we make use of the ED method as an impurity solver, 
where the fictitious temperature\cite{Caffarel} $\tilde{\beta}$ allows 
us to solve the self-consistent equation
$G_{loc}=G_{imp}$ numerically.
In this paper, we use the fictitious temperature $\tilde{\beta}(\geq 50)$
and the number of sites for the impurity model is set as $N=6$
to converge the DMFT iterations.
We note that a careful scaling analysis for the fictitious temperature and 
the number of sites is needed only when the system is near the metal-insulator
transition points.
To discuss the stability of the Fermi liquid state, we define 
the quasi-particle weight for $\alpha$th band as
\begin{eqnarray}
Z_\alpha =\left.\left(
\frac{ 1-{\rm Im} \Sigma_\alpha(\tilde{\omega}_n) }
{ \tilde{\omega}_n}\right)^{-1}\right|_{n=0},
\end{eqnarray}
where $\tilde{\omega}_n[=(2n+1)\pi/\tilde{\beta} ]$ is the Matsubara frequency.

\begin{figure}[htb]
\begin{center}
\includegraphics[width=7cm]{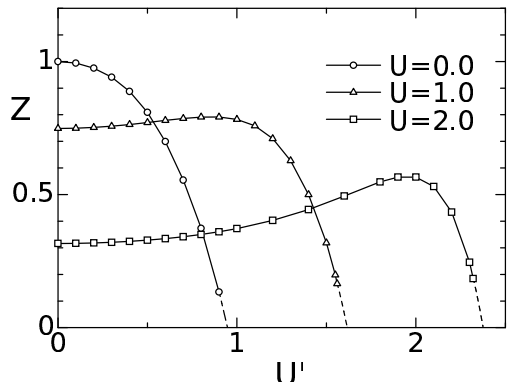}
\end{center}
\vskip -4mm
\caption{The quasi-particle weight $Z$ as a function of the interorbital
 Coulomb 
interaction $U'$. The data are obtained by the ED $(N=6)$ within DMFT.}
\label{fig:Z1}
\end{figure}

\begin{figure}[htb]
\begin{center}
\includegraphics[width=7cm]{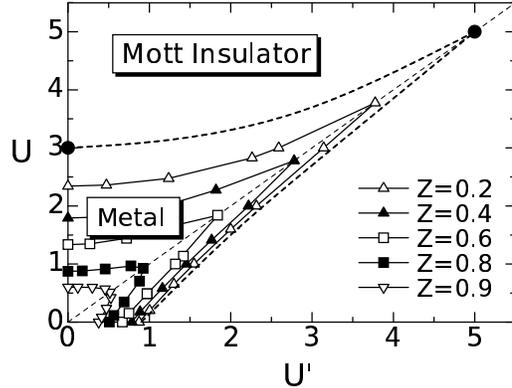}
\end{center}
\vskip -4mm
\caption{The zero-temperature phase diagram for $J=0$.
The contour plot of the quasi-particle weight $Z$ is shown:
the bold-dashed line represents the phase boundary of the metal-insulator 
transition, which is obtained  by estimating 
the values of $U$ and $U'$ that give $Z=0$.
The solid circles are the transition points obtained by the 
linearized DMFT.\cite{LDMFT}
}
\label{fig:phasephase}
\end{figure}

In Fig. \ref{fig:Z1},
the quasi-particle weight 
calculated 
is shown 
as a function of the interorbital interaction 
$U'$ for several different values  of intraorbital interaction $U$.
For $U=0$, $Z$ decreases monotonically with 
increasing $U'$, and a metal-insulator transition occurs around
 $U'_c\sim 0.9$.  On the other hand,
 when $U\neq 0$, there appears nonmonotonic behavior
 in $Z$; it once increases with the increase of $U'$,
 has the maximum value  in the vicinity of $U'\sim U$, 
and finally vanishes at the Mott transition point. 
It is somehow unexpected that 
the maximum structure appears around $U\sim U'$, which is more enhanced
for larger $U$ and $U'$.  We will see below that this 
is related to enhanced orbital fluctuations.
By repeating similar calculations systematically, we end up with the 
phase diagram at zero temperature (Fig.  \ref{fig:phasephase}), where
the contour plot of the quasi-particle weight is shown explicitly.
At $U'=0$, where the system is reduced to the single-orbital Hubbard model, 
we find that the intraorbital Coulomb interaction $U$ triggers 
the metal-insulator transition at $U_c=2.9\sim3.0$.
This critical value obtained by the ED 
of a small system $(N=6)$ is in good agreement with other
 numerical calculations such as  
the numerical renormalization group $(U_c=2.94)$,\cite{BullaNRG,Zitzler}
the ED $(U_c=2.93)$,\cite{OnoED} 
the linearized DMFT $(U_c=3)$,\cite{LDMFT,2site}
the dynamical density-matrix renormalization group 
$(U_c=3.07)$.\cite{Nishimoto,Uhrig}

There are some notable features in the phase diagram.
First,  the value of $Z$ is not so sensitive to
$U'$ for a given $U$ ($>U'$), except for the 
region $U \sim U'$. In particular, the phase boundary indicated by 
the dashed line in the upper side of the figure
 is almost flat for the small $U$ region. 
The second  point is that when $U \sim U'$ the metallic 
phase is stabilized  up to fairly large Coulomb interactions, and it
 becomes unstable,
once the parameters are away from the condition $U=U'$.  
The latter tendency is more conspicuous in the regime of strong correlations.
\cite{Koga,Bunemann:Gutzwiller}

\subsection{finite-temperature properties}

\begin{figure}[htb]
\begin{center}
\includegraphics[width=7cm]{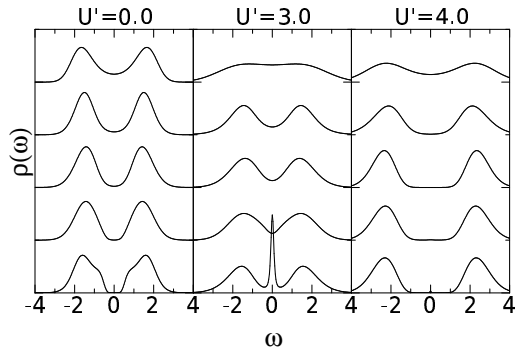}
\end{center}
\vskip -4mm
\caption{The DOS for the two-orbital Hubbard model $(U=3.0)$. 
The data are for the inverse 
temperature $T^{-1}=1, 2, 4, 8$ and $16$ from the top to the bottom.}
\label{fig:dosdos}
\end{figure}

To observe such characteristic properties around $U = U'$
in more detail, we compute the physical quantities at finite temperatures
by exploiting the QMC method as an impurity solver.\cite{Hirsch}
In Fig. \ref{fig:dosdos}, we show the DOS deduced by 
applying the maximum entropy method (MEM) \cite{MEM1,MEM2,MEM3} 
to the Monte Carlo data.\cite{KogaSN}
In the case  $(U, U')=(3.0, 0.0)$, the system is in the insulating phase 
close to the transition point (see Fig.\ref{fig:phasephase}). 
Nevertheless we can clearly observe the formation of the Hubbard gap
with the decrease of temperature $T$.
In the presence of $U'$, the sharp quasi-particle peak is 
developed around 
the Fermi level at low temperatures (see the case of
 $(U, U')=(3.0, 3.0)$).
This implies that orbital fluctuations induced by $U'$ drive the 
system to the metallic phase.
Further increase in the interorbital interaction suppresses  
 spin fluctuations, leading 
the system to another type of the Mott insulator in the 
region of $U'>U$.

\begin{figure}[htb]
\begin{center}
\includegraphics[width=7cm]{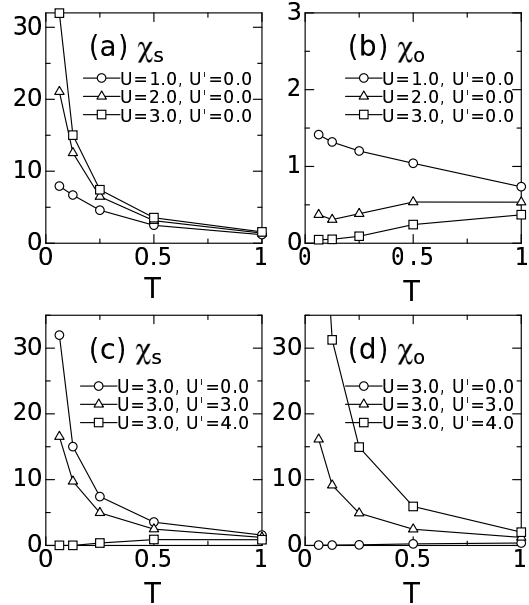}
\end{center}
\vskip -4mm
\caption{The local spin and orbital susceptibilities 
as a function of the temperature.}
\label{fig:chi}
\end{figure}

To characterize the nature of the spin and orbital 
fluctuations around  the Mott transition,
we investigate the temperature dependence of the 
local spin and orbital susceptibilities,
which are defined as,
\begin{eqnarray}
\chi_s&=&\int_0^\beta {\rm d}\tau\langle
\left\{n_\uparrow(0)-n_\downarrow(0)\right\}
\left\{n_\uparrow(\tau)-n_\downarrow(\tau)\right\}\rangle\nonumber\\
\chi_o&=&\int_0^\beta {\rm d}\tau\langle
\left\{n_1(0)-n_2(0)\right\}
\left\{n_1(\tau)-n_2(\tau)\right\}\rangle,
\end{eqnarray}  						  
where $\beta=T^{-1}$, $n_\sigma=\sum_\alpha n_{\alpha,\sigma}$, 
$n_\alpha=\sum_\sigma n_{\alpha, \sigma}$ and $\tau$ is imaginary time.
We show the results obtained by QMC simulations
within DMFT in Fig. \ref{fig:chi}. 
Let us first look at Figs. \ref{fig:chi} (a) and (b) for  $U'=0$ 
(equivalent to the single-orbital model). Since
the introduction of the intraorbital interaction $U$ makes
the quasi-particle peak narrower, it increases 
the spin susceptibility $\chi_s$ at low temperatures. On 
the other hand, the formation of the Hubbard gap 
suppresses not only the charge susceptibility
but also the orbital susceptibility $\chi_o$.  
As seen from Figs. \ref{fig:chi} (c) and (d),
 quite different behavior appears in the susceptibilities, 
when   the interorbital interaction $U'$ is increased. 
Namely, the spin susceptibility is suppressed, while
the orbital susceptibility is enhanced
at low temperatures. This tendency  holds 
 for larger $U'$ beyond the condition $U' \sim U$.
Therefore in the metallic phase close to the Mott insulator
in the region of  $U>U' (U<U')$, spin (orbital) fluctuations 
are enhanced  whereas orbital (spin) fluctuations are suppressed with 
the decrease of  the temperature.
These analyses clarify 
why the metallic phase is particularly stable along the 
line $U=U'$.  Around this line, spin and orbital fluctuations
are almost equally enhanced, and this subtle balance is efficient
to stabilize the metallic phase. 
When the ratio of interactions deviates from this condition, the 
system prefers either of the two Mott insulating phases.

\subsection{phase diagram at finite temperatures}

\begin{figure}[htb]
\begin{center}
\includegraphics[width=7cm]{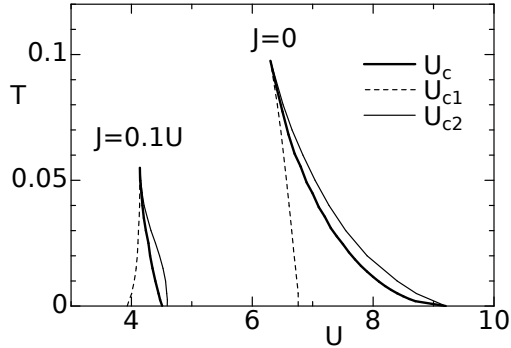}
\end{center}
\caption{
The finite temperature phase diagram for the two-orbital Hubbard 
model for $J=0$ and $J=0.1U$ under the condition $U=U'+2J$.
}\label{hund}
\end{figure}

QMC simulations are not powerful enough to determine the phase 
diagram at finite temperatures. To overcome this difficulty, we 
make use of a complementary method, SFA,\cite{InabaB} 
which allows us to discuss the Mott transition at 
finite temperatures.
The obtained phase diagram is shown in Fig. \ref{hund},
where not only the case of $J=0$ but also 
$J=0.1U$ are studied under the condition $U=U'+2J$.

It should be noted that the Mott transition at finite temperatures is 
of first order, similarly to the single-orbital Hubbard case.\cite{Georges}
The critical value for the transition point, $U_c$, is 
determined by comparing the grand potential for each phase. 
We find the region in the phase diagram 
where the metallic and Mott insulating phases coexist.
Starting from the metallic phase at low temperatures,
the increase of the Coulomb interaction 
 triggers the Mott transition to the insulating phase 
 at $U_{c2}$, where we can observe the discontinuity 
in the quasi-particle weight $Z$.
On the other hand, the Mott transition occurs at $U_{c1}$ 
when the interaction decreases. 
The phase boundaries $U_c$, $U_{c1}$ and $U_{c2}$ merge to 
the critical temperature $T_c$, where the second order transition occurs.

Note that upon introducing the Hund coupling $J$, the phase boundaries
are shifted to the weak-interaction region, and 
therefore the metallic state gets unstable for large $U$. 
Also, the coexistence region surrounded by 
the first order transitions shrinks as $J$ increases. This tendency
reflects the fact that the metallic state is stabilized by 
enhanced orbital fluctuations  around  $U=U'$: the Hund 
coupling suppresses such orbital fluctuations, and 
stabilizes the Mott-insulating phase.
 Another remarkable point is that the Mott transition becomes of
first-order even at zero temperature in the presence of
$J$,\cite{Bunemann:Gutzwiller,Ono03,Pruschke} 
since the subtle balance realized at $T=0$ in the 
case of $U=U'$ is not kept anymore for finite $J$.
There is another claim that the second order transition could be 
possible for certain 
choices of the parameters at $T=0$,\cite{Pruschke} 
 so that more detailed discussions may be
necessary to draw a definite conclusion on this problem.

\section{Orbital-selective Mott transitions}\label{sec4}

In the previous section, we investigated the two-orbital Hubbard model 
with same bandwidths, and found that the metallic state is 
stabilized up to fairly large Coulomb interactions around $U = U'$,
which is shown to be caused by the enhanced spin and orbital fluctuations.
We now wish to see what will happen if we consider 
the system with different bandwidths,
which may be important in real materials such as
$\rm Ca_{2-x}Sr_xRuO_4$\cite{Nakatsuji1} 
and $\rm La_{n+1}Ni_nO_{3n+1}$.\cite{Kobayashi96,LaNiO}
In the following, we will demonstrate that the enhanced spin and
orbital fluctuations again play a key role in controlling the 
nature of the Mott transitions even for the system with different 
bandwidths.\cite{KogaLett,KogaB,Inaba}

\subsection{zero-temperature properties}

\begin{figure}[htb]
\begin{center}
\includegraphics[width=6cm]{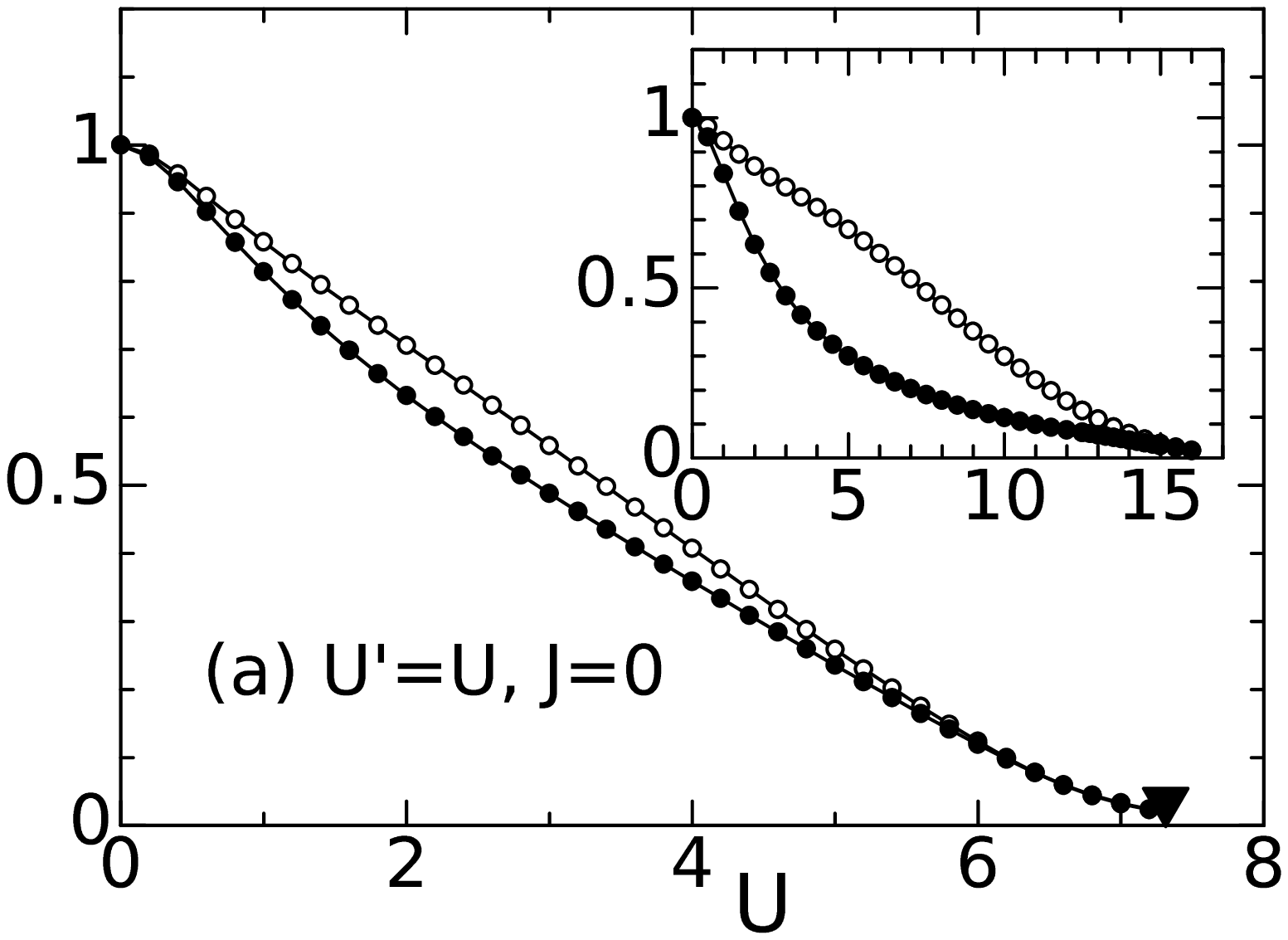}
\includegraphics[width=6cm]{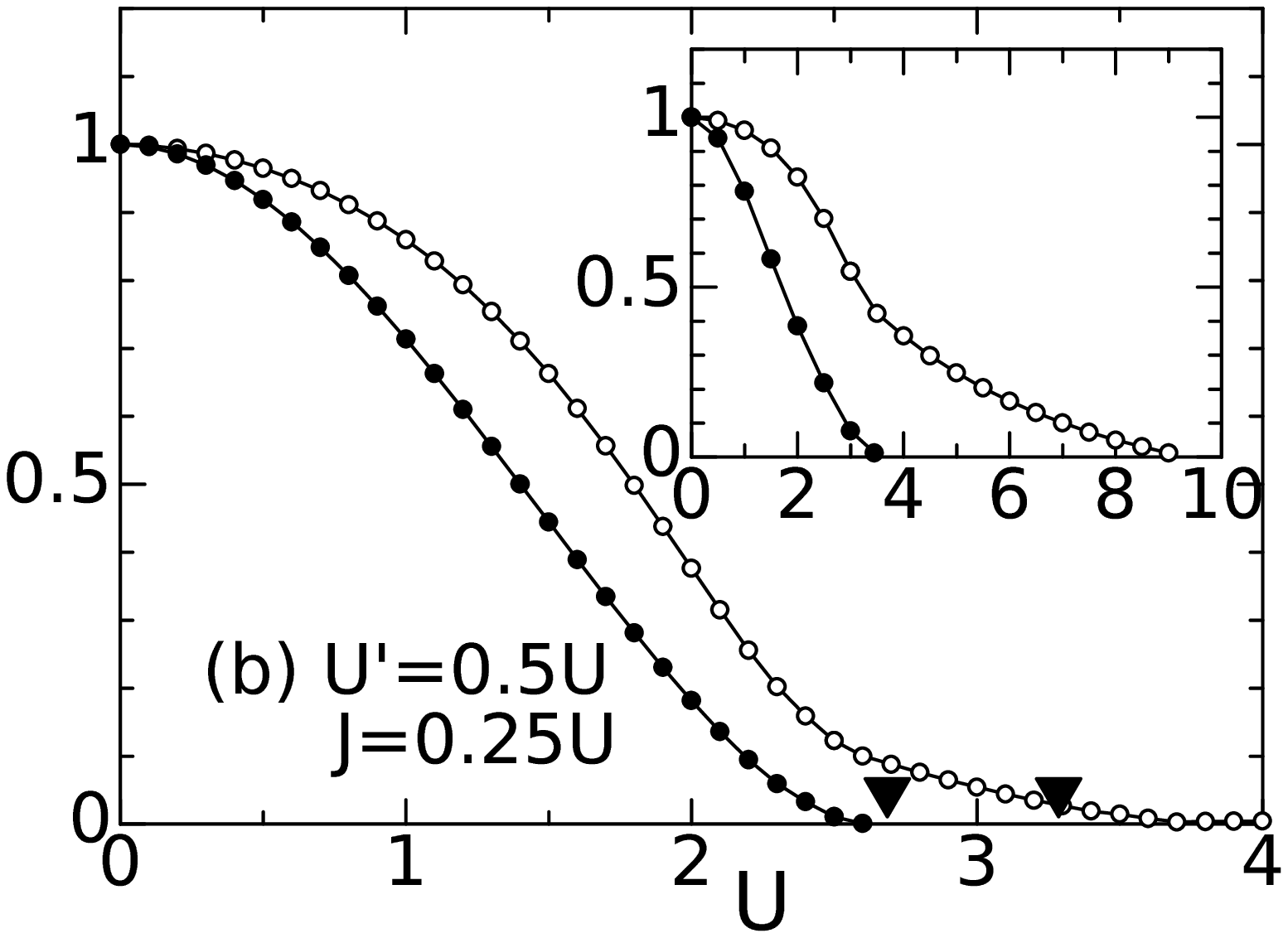}
\end{center}
\caption{The quasi-particle weights $Z_1$ and $Z_2$ at half filling
as a function of $U$ for $D_1=1.0$ and $D_2=2.0$: (a) $U'/U=1.0$ ($J=0$)  and 
(b) $U'/U=0.5$ ($J/U=0.25$). 
Open (closed) circles represent the results for orbital $\alpha=1(2)$
obtained by combining DMFT the ED as an impurity solver 
for the $N=6$ small cluster.
Solid triangles represent the Mott-transition points
obtained by the two-site DMFT method. Insets show the same plots 
for $D_1=1.0$ and $D_2=5.0$.
}
\label{fig:Z}       
\end{figure}

Let us start with the quasi-particle weight calculated
by DMFT with the ED method \cite{KogaLett} and  see the stability 
of the metallic phase at zero temperature. Here,
we include the Hund coupling explicitly under the constraint $U=U'+2J$.
The quasi-particle weights obtained with fixed ratios $U'/U$ and $J/U$
are shown in Fig. \ref{fig:Z} for half-filled bands.
We first study the case of $U=U'$ and $J=0$
 with bandwidths $D_1=1.0$ and $D_2=2.0$ [Fig. \ref{fig:Z} (a)].
When the Coulomb interaction is switched on, the 
quasi-particle weights $Z_1$ and $Z_2$ decrease
from unity in slightly
different manners reflecting the different bandwidths.
A strong reduction in  the quasi-particle weight 
appears initially for the narrower band. 
However, as the system approaches the Mott transition, 
the quasi-particle weights merge again, exhibiting very similar
$ U $-dependence, and eventually vanish at 
the same critical value.  This behavior is explained as follows.
For small interactions,
the quasi-particle weight depends on the effective Coulomb interactions 
$U/D_\alpha$ which are different for two bands of different width
$D_{\alpha}$, and give distinct behavior of $Z_1 $ and $Z_2$.
However, in the vicinity of the Mott transition, the
effect of the bare bandwidth is diminished due to the
strong renormalization of the effective quasi-particle bandwidth,
allowing $Z_1$ and $Z_2$ to vanish together
\cite{Gutzwiller,Brinkman}. 

The introduction of a finite Hund coupling $J$ makes $ U \neq U' $,
and causes qualitatively different behavior [Fig. \ref{fig:Z} (b)].
As $ U $ increases with the fixed ratio $ U'/U=0.5 $, 
the quasi-particle weights decrease differently and vanish at different
critical points: $U_{c1}\approx 2.6$ for $ Z_1$ and $U_{c2} \approx
3.5 $ for $Z_2 $. We thus have an intermediate phase with one
orbital localized and the other itinerant, which may be
referred to as the intermediate phase. The analogous 
behavior is observed for
other choices of the  
bandwidths, if $J$ takes a finite value [see the 
inset of Fig. \ref{fig:Z} (b)].
These results certainly suggest the existence of
the OSMT with $ U_{c2} > U_{c1} $.

\begin{figure}[htb]
\begin{center}
\includegraphics[width=7cm]{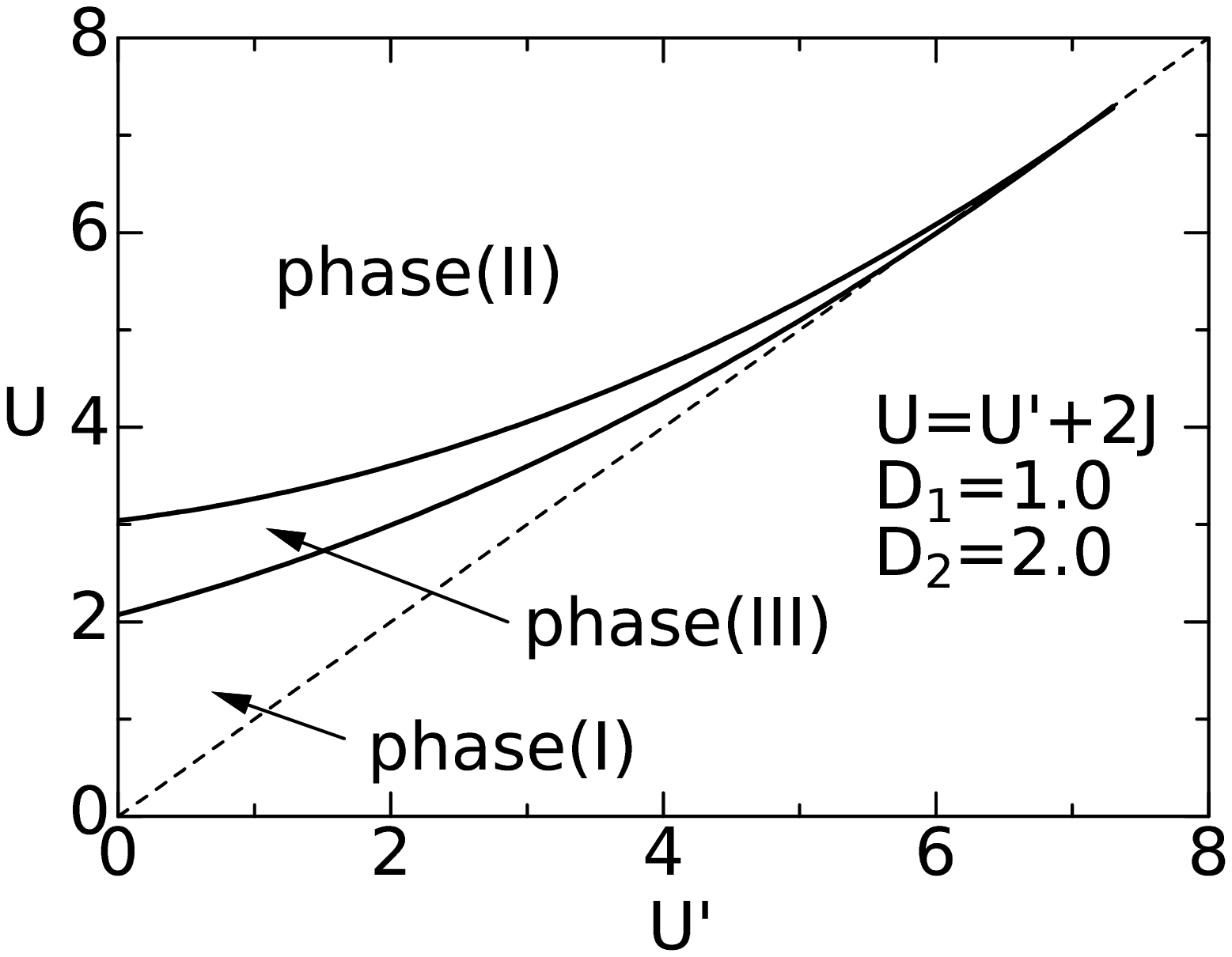}
\end{center}
\caption{The zero-temperature phase diagram for two-orbital Hubbard model
for  $D_1=1$ and $D_2=2$.  In the phase (I) [phase (II)], 
both bands are in  metallic (insulating) state.
The phase (III) is induced
by the OSMT, where 
the metallic state coexists with the Mott insulating state.
Since we consider the ferromagnetic Hund coupling, $J>0$,
the relevant region in the diagram is $U>U'$.
}
\label{fig:phase}       
\end{figure}

We have repeated  similar DMFT calculations for various 
choices of the parameters to  determine 
the ground-state phase diagram, which is shown in Fig. \ref{fig:phase}.
The phase diagram has some remarkable features.
First, the metallic phase (I) is stabilized up to fairly
large Coulomb interaction $U$ when $U \to U'$ (small $J$).
Here the Mott transitions merge to a single transition. 
As mentioned above, this behavior reflects the high symmetry
in the case of 
$U=U' $ ($J=0$) with six degenerate two-electron onsite 
configurations: four spin configurations with one electron in each orbital
and two spin singlets with both electrons in one of the two orbitals.  
Away from the symmetric limit, i.e.  $ U > U' $ ($ 2J = U - U' $) orbital
fluctuations are suppressed and the spin 
sector is reduced by the Hund coupling to three onsite spin triplet
components as the lowest multiplet for two-electron states.
In this case, we encounter two types of the Mott transitions
having different critical points.
In between the two transitions we find the intermediate OSM phase (III)
with one band localized and the other itinerant. 
Within our DMFT scheme we have confirmed that various choices 
of bandwidths give rise to the qualitatively same structure of the
phase diagram as shown in Fig. \ref{fig:phase} (see also the discussions 
in Summary).

\subsection{finite-temperature properties}

To clarify how enhanced orbital fluctuations 
around  $U=U'$ affect the nature of Mott transitions,
\begin{figure}[htb]
\begin{center}
\includegraphics[width=7cm]{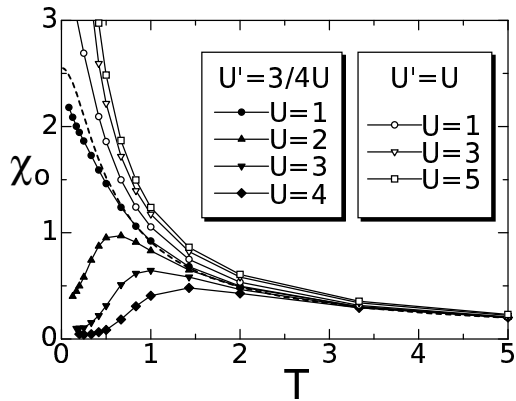}
\end{center}       
\caption{ The
orbital susceptibility as a function of $T$ for $V=0$. 
Open (solid) symbols represent the results in the case $U=U'$ and $J=0$ 
($U'/U=3/4$ and $J/U=1/8$) and dashed lines those for the 
non-interacting case.
}
\label{fig:chi-o}
\end{figure}
let us study the temperature-dependent 
orbital susceptibility, which is shown in Fig. \ref{fig:chi-o}.
Here, we have used the new algorithm of the QMC 
simulations proposed by Sakai {\it et al.}\cite{Sakai} 
to correctly take into account the effect of Hund coupling
including the exchange and the pair hopping terms.\cite{KogaB}

We can see several characteristic properties in Fig. \ref{fig:chi-o}.
In the non-interacting system,
the orbital susceptibility increases with decreasing temperature,
and reaches a constant value at zero temperature.
When the interactions are turned on ($U'/U=3/4$ and $J/U=1/8$), 
the orbital susceptibility is suppressed
at low temperatures, which implies that electrons 
in each band become independent. Eventually 
for $U \ge U_{c1} \sim 3$, one of the orbitals is localized, 
so that orbital fluctuations are suppressed completely, giving 
$\chi_o=0$ at $T=0$.
On the other hand, quite different behavior appears 
around $U'=U$. In this case, the
orbital susceptibility is increased with the increase of interactions
even at low temperatures. 
Comparing this result with  the  phase diagram in 
Fig. \ref{fig:phase},
we can say that the enhanced orbital fluctuations are 
relevant for stabilizing the metallic phase in the
strong correlation regime. While such behavior has
been observed for models with
two equivalent orbitals (previous section), it appears even in 
the systems with
nonequivalent bands.\cite{KogaSN}

\begin{figure}[htb]
\begin{center}
\includegraphics[width=6cm]{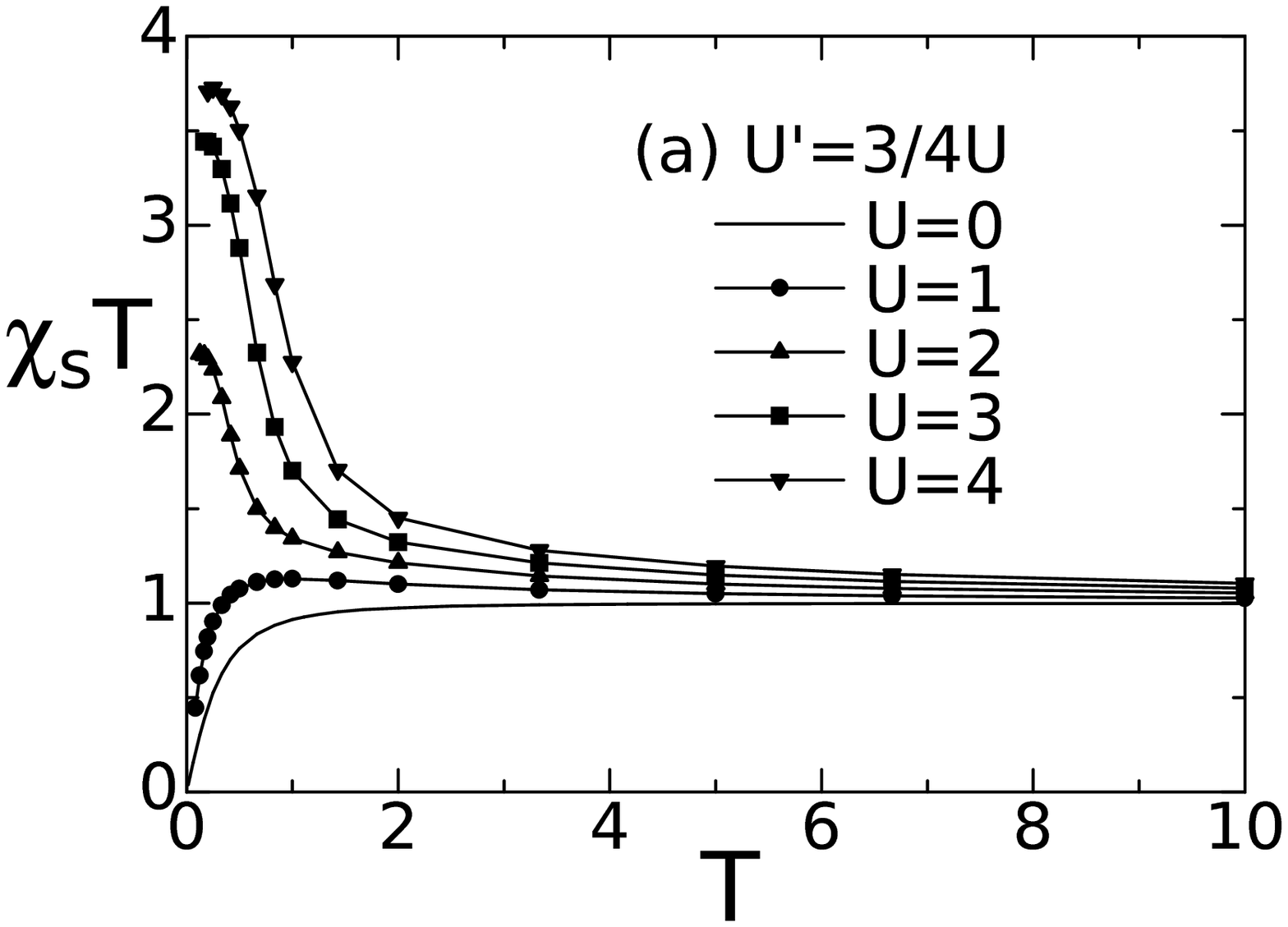}
\includegraphics[width=6cm]{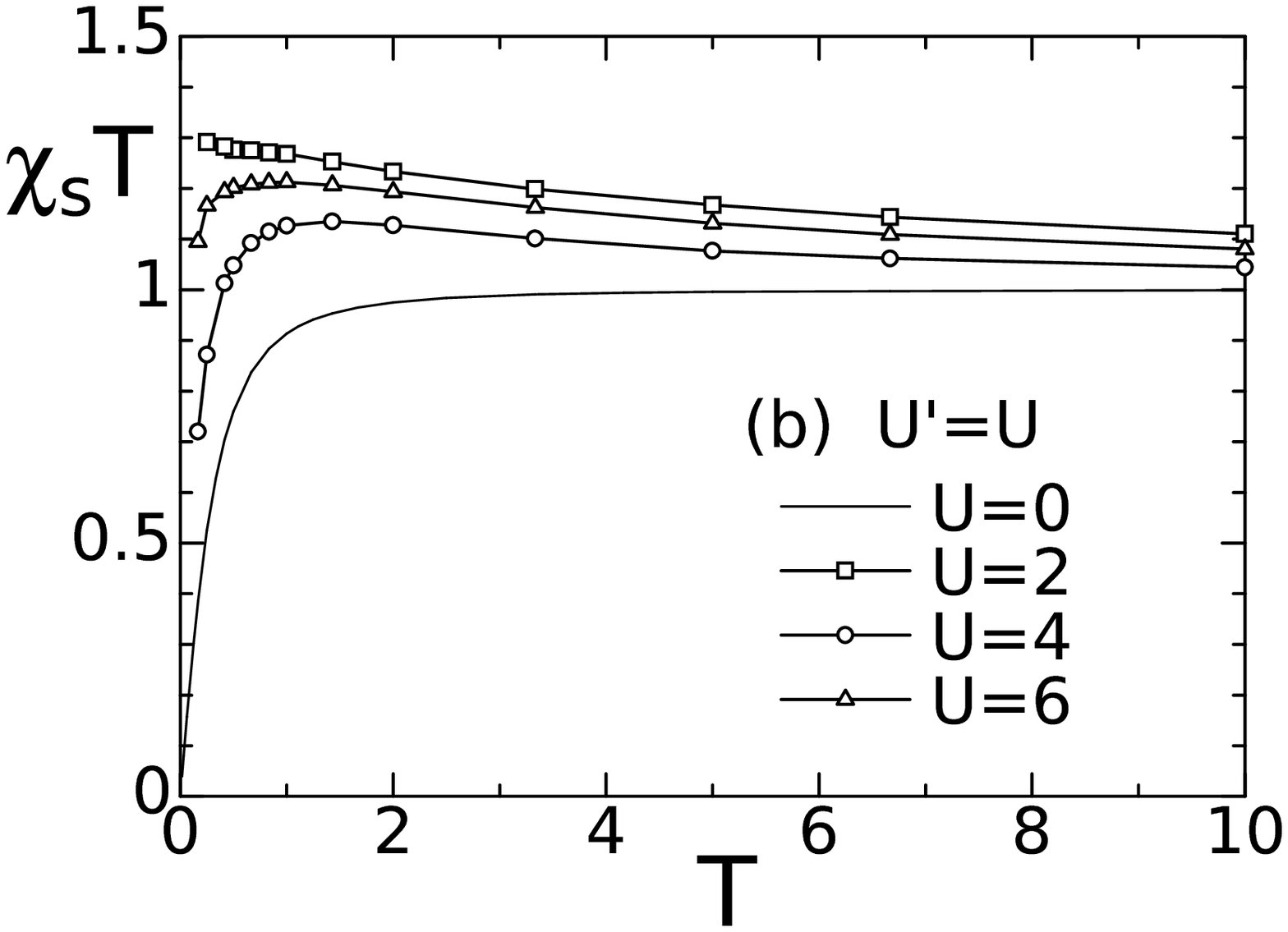}
\end{center}       
\caption{
(a) The effective Curie constant
 $\chi_s T$ as a function of  $T$ 
for $U'/U=3/4$ and $J/U=1/8$; (b) the results for $U'=U$ and $J=0$.
}
\label{fig:chit-s}
\end{figure}
In Fig. \ref{fig:chit-s},  
the effective Curie constant  $\chi_s T$  is shown as a function 
of the temperature.
We first look at the case of $U'/U=3/4$ and $J/U=1/8$.
At high temperatures, all the spin configurations are equally 
populated, so that the effective Curie constant has the value
$1/2$ for each orbital in our units, giving $\chi_s T\sim 1$.
For weak electron correlations
($U=1$), the system is in the metallic phase,
so that the Pauli paramagnetic behavior
 appears, resulting in  $\chi_s T \rightarrow 0$  
as $T \rightarrow 0$. We can see that the increase of the interactions
enhances the spin susceptibility at low temperatures, as a result of
the progressive trend to localize the electrons. 
The effective Curie constant is $\chi_sT=2$ when a free spin is realized
in each orbital, while $\chi_sT=8/3$ 
when a triplet $S=1$ state is realized due to the Hund coupling.
It is seen that the Curie constant increases beyond these values 
with the increase of the interactions ($U=3, 4$).
This means that ferromagnetic correlations due to 
the Hund coupling appear here. 

For $U'=U$, we can confirm that not only orbital but also 
spin fluctuations are enhanced in the 
presence of the interactions, see Fig. \ref{fig:chit-s} (b).
Accordingly, both  spin and orbital susceptibilities
increase at low temperatures, forming  heavy-fermion states
as far as the system is in the metallic phase.  Note that 
for $U=6$, at which 
the system is  close to the  Mott
transition point, the spin susceptibility is enhanced with the effective
Curie constant $\chi_sT \sim 4/3$ down to very low temperatures
[Fig. \ref{fig:chit-s} (b)].
The value of 4/3 originates from 
two additional configurations of doubly-occupied orbital besides
 four magnetic configurations, which are all degenerate
 at the metal-insulator 
transition point. Although not clearly seen
in the temperature range shown, the Curie constant  $\chi_sT$ should vanish at
zero temperature for $U=U'=6$, since the system is still in 
the metallic phase, as seen from Fig. \ref{fig:phase}.

\begin{figure}[htb]
\begin{center}
\includegraphics[width=7cm]{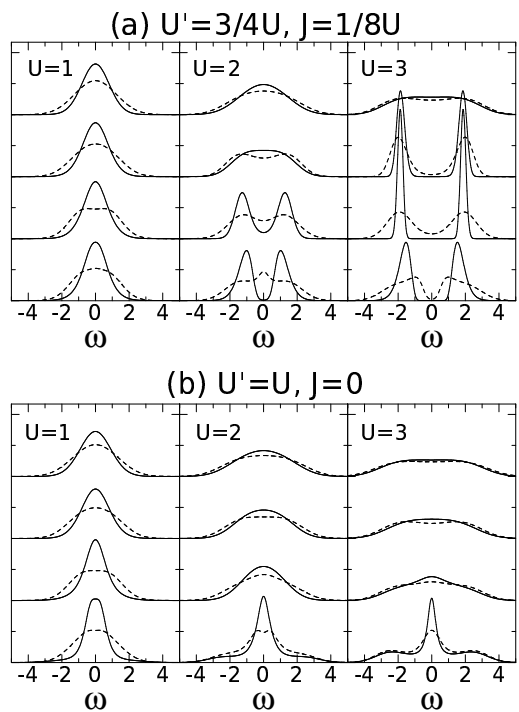}
\end{center}       
\caption{
Solid (dashed) lines represent the DOS
for the orbital $\alpha=1$ ($\alpha=2$)
when $(D_1, D_2)=(1.0, 2.0)$. The data are for the temperatures $T=2, 1,
 1/2$ and $1/6$ from the top to the bottom.
}
\label{fig:dos0}
\end{figure}

To see how the above spin and orbital characteristics affect
one-electron properties, we show 
the DOS  for each orbital in Fig. \ref{fig:dos0},
which is computed by the MEM.\cite{MEM1,MEM2,MEM3}
When the interactions increase along the line  $U'/U=3/4$ and $J/U=1/8$,
 the OSMT should occur. Such tendency indeed 
appears at low temperatures in Fig. \ref{fig:dos0}(a).
Although both orbitals are in metallic states down to
low temperatures ($T=1/6$) for $U=1$,  the OSMT
seems to occur for $U=2$; one of the bands develops the Mott Hubbard
gap, while the other band still remains metallic.
At a first glance, this result seems slightly different from
 the ground-state phase diagram 
shown in Fig. \ref{fig:phase}, where the system is in 
the phase (I) even at  $U=2$.  
However, this deviation is naturally understood
if we take into account the fact that for $U=2$, the 
narrower band is already in a highly correlated  
 metallic state, so that the sharp quasi-particle peak immediately 
disappears as the temperature increases. This explains the behavior 
observed in the DOS at $T=1/6$.  For $U=3$, both 
bands are insulating at $T=1/6$ (the system
is near  the boundary between the phases (II) and (III) at 
$T=0$).
For $U'=U$, the qualitatively different 
behavior appears  in Fig. \ref{fig:dos0}.
 In this case, quasi-particle peaks are developed in 
both bands as the interactions increase, and the system 
 still remains metallic even at $U=U'=3$.
  As mentioned above, all these features, which are contrasted 
to  the case of $U' \neq U$, 
are caused by equally enhanced spin and orbital fluctuations 
around $U=U'$.

\subsection{phase diagram at finite temperatures}

\begin{figure}[htb]
\begin{center}
\includegraphics[width=7cm]{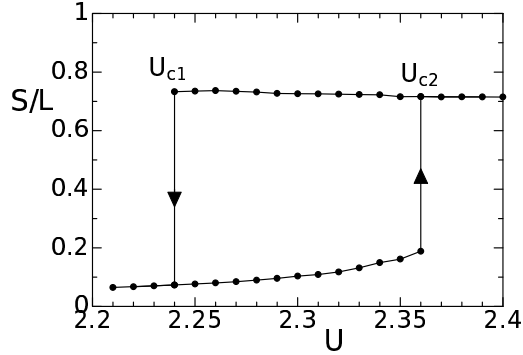}
\end{center}
\caption{
The entropy as a function of  $U$ at $T=0.002, J=0.25U$.
}
\label{fig:local}
\end{figure}

\begin{figure}[htb]
\begin{center}
\includegraphics[width=7cm]{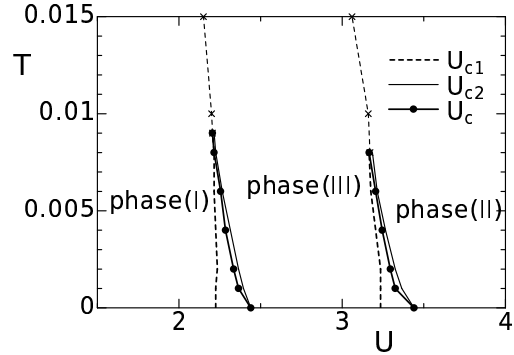}
\end{center}
\caption{The finite-temperature  phase diagram for $J=0.25U$. 
There are two coexistence phases, which correspond to the 
triangular-shaped regions: the metallic phase and 
the intermediate phase coexist in the left region  and 
the intermediate phase and insulating phase coexist in the 
right region.}
\label{fig:phase_J25}
\end{figure}

\begin{figure}[htb]
\begin{center}
\includegraphics[width=6cm]{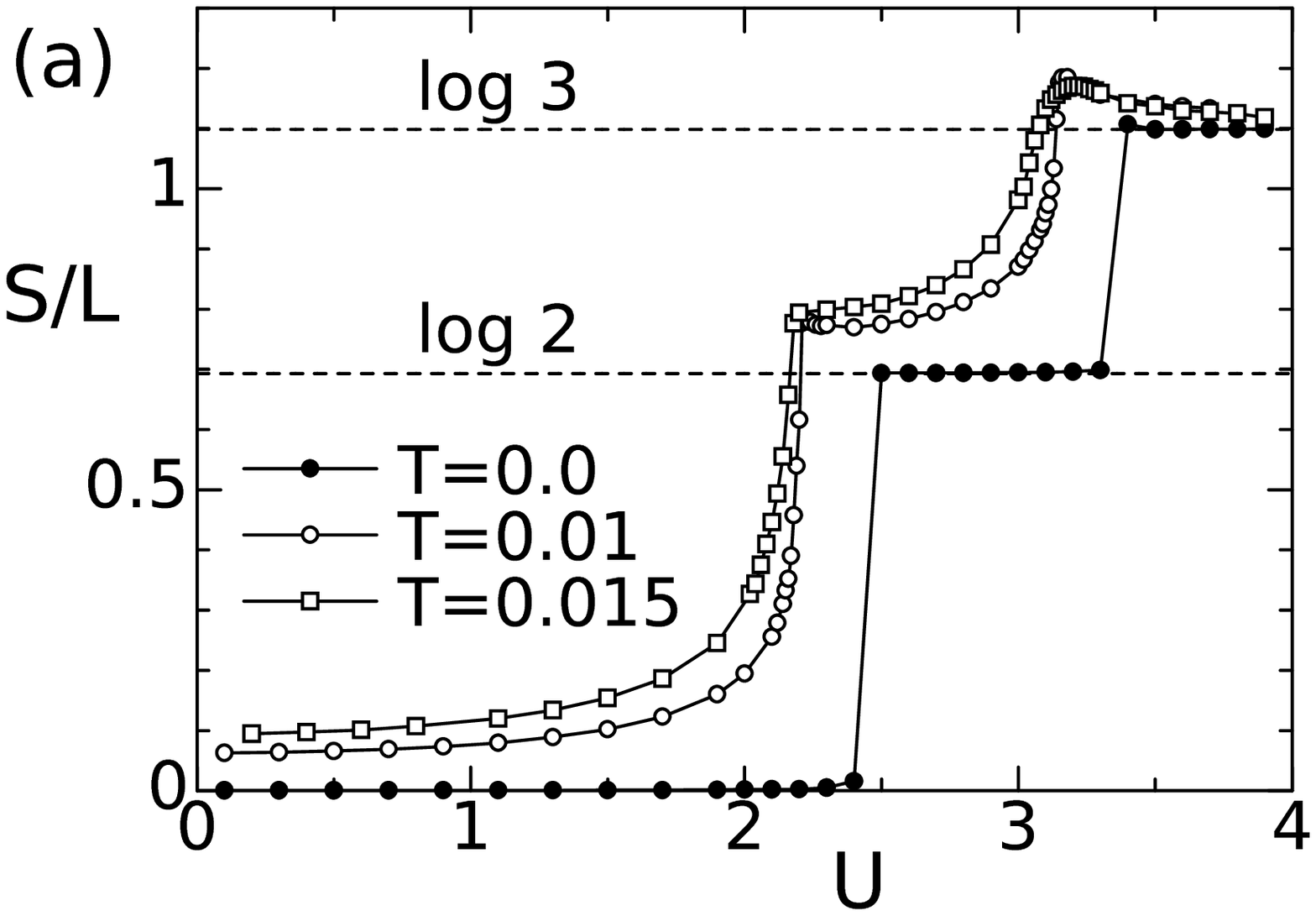}
\includegraphics[width=6cm]{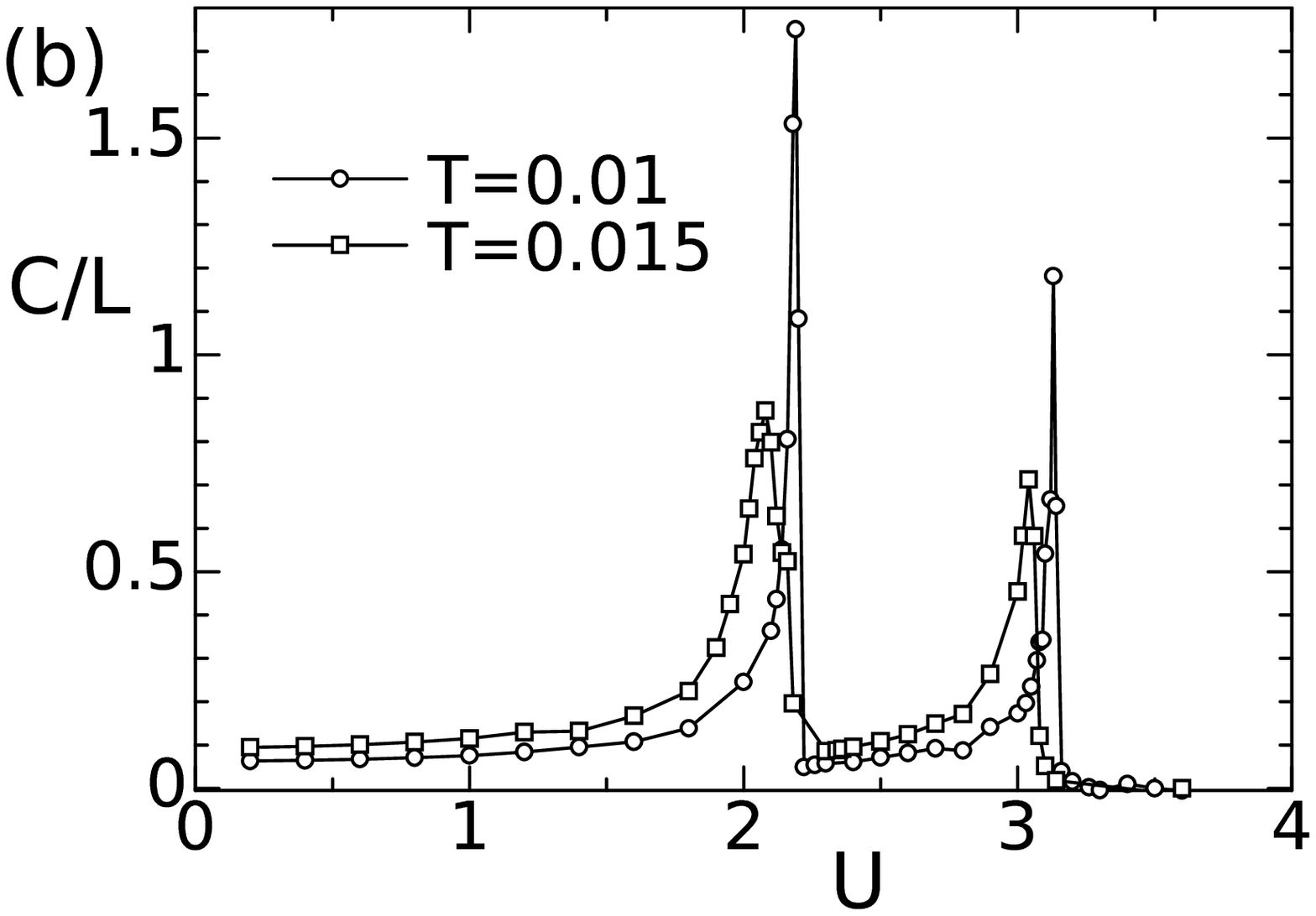}
\end{center}
\caption{(a) Entropy $S/L$ and (b) Specific heat $C/L$
as a function of $U$ in the crossover region, $J=0.25U$.
}\label{fig:hcap}
\end{figure}

Having studied the spin and orbital properties,
we now obtain the phase diagram at finite temperatures
in the general case $U\neq U'$ and $J\neq 0$.
Since each Mott transition at zero
temperature is similar to that for the single-orbital Hubbard model,
\cite{Georges} we naturally expect that the transition should be 
of first order at finite temperatures 
around each critical point \cite{LiebschFiniteT}.
In fact, we find the hysteresis in the physical quantities. 
For example, we show  the entropy per site in Fig. \ref{fig:local}.
At $T=0.002$,  when $U$ increases,
the metallic state (I) disappears around $U_{c2}\sim 2.36$ 
where the first-order Mott transition  
occurs to the intermediate phase (III), which is accompanied by
 the jump in the curve of entropy.
On the other hand, as $U$ decreases, the intermediate phase (III)
is stabilized down to $U_{c1}\sim 2.24$.
The first-order critical point $U_c\sim 2.33$ for $T=0.002$ is estimated
by comparing the grand potential for each phase.\cite{Inaba}
The phase diagram thus obtained by SFA is shown in Fig. \ref{fig:phase_J25}.
It is seen that the two  coexistence regions 
appear around $U\sim 2.4$ and $U\sim 3.3$.
The phase boundaries $U_{c1}$, $U_{c2}$ and $U_c$ merge 
at the critical temperature $T_c$ for each transition.
 We note that  similar phase diagram 
was obtained by Liebsch by means of DMFT with 
the ED method.\cite{LiebschFiniteT}  Our SFA treatment 
elucidates further interesting 
properties such as the crossover behavior among the competing 
phases (I), (II) and (III).  To see this point more clearly,
we show the detailed results for
 the entropy and the specific heat in Fig. \ref{fig:hcap}.
There exists a double-step structure in the curve of the entropy,
which is similar to that found at $T=0$, 
where the residual entropy 
$S/L=0, \log 2$ and $\log 3$ appears in the phase (I), (III) and (II),
respectively. Such anomalies
 are observed more clearly in  the specific heat.  It is remarkable  
that the crossover behavior among three phases is clearly seen
even at high temperatures. Therefore, 
the intermediate phase (III) is  well defined even at higher 
temperatures above the critical temperatures.

\section{Effect of hybridization between distinct orbitals}\label{sec5}

In the present treatment of the model 
with DMFT, the intermediate phase (III) is unstable 
against certain perturbations. There are several 
mechanisms that can stabilize this phase. A possible mechanism,
which may play an important role in real materials,
is the hybridization between the two distinct orbitals.
The effect of the  hybridization is indeed important, e.g.  
for the compound $\rm Ca_{2-x}Sr_x Ru O_4$, \cite{Nakatsuji1}
where the hybridization between $\{\alpha, \beta\}$ and $\gamma$ orbitals 
is induced by the tilting of RuO$_6$ octahedra in the 
region of  $\rm Ca$-doping $0.2<x<0.5$\cite{tilting}. An interesting point 
is that the hybridization effect could be closely 
related to the reported heavy 
fermion behavior.\cite{Nakatsuji,Nakatsuji1}
The above interesting aspect naturally motivates us to study  the 
hybridization effect
between the localized and itinerant electrons in the 
intermediate phase (III).   
Here, we take into account the effect of hybridization in each phase
to study the instability of the OSMT.\cite{KogaB,Medici05}

\begin{figure}[htb]
\begin{center}
\includegraphics[width=7cm]{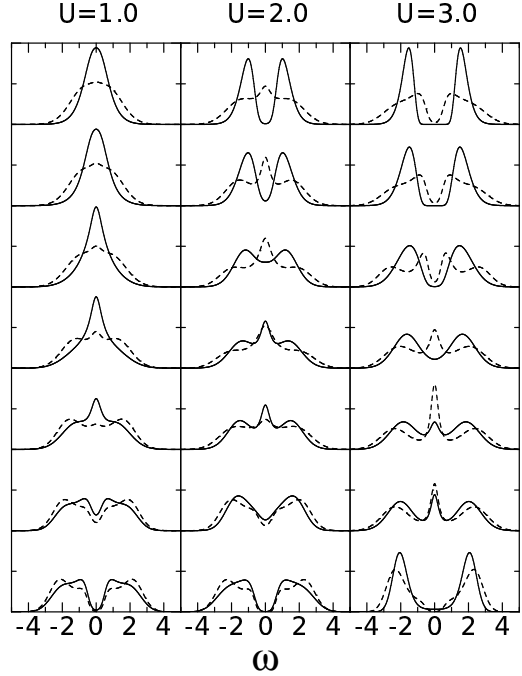}
\end{center}       
\caption{Solid (dashed) lines represent the DOS
for the orbital $\alpha=1$ ($\alpha=2$)
when $(D_1, D_2)=(1.0, 2.0)$ at $T=1/6$
with  the  fixed  parameters of $U'/U=3/4$ and $J/U=1/8$.
The data are plotted for $V=0.0, 0.25, 0.5, 0.75, 1.0, 1.25$ 
and $1.5$ from top to bottom.
}
\label{fig:dos-V}
\end{figure}

We study the general case with
$U'\neq U$ and $J\neq 0$ in the presence of the hybridization
$V$. In Fig. \ref{fig:dos-V}, we show the 
 DOS calculated by QMC and MEM for several different values of  $V$.
We begin with the case of weak interaction, $U=1$, where
the metallic states are realized in both orbitals at $V=0$.
Although the introduction of small $V$ does not change
 the ground state properties, further increase in $V$ splits
the DOS, signaling the formation of the 
band insulator,  where all kinds of excitations have the gap.
On the other hand, different behavior appears when the
interactions are increased up to $U=2$ and 3. In these cases,
 the system at $V=0$ is in  the intermediate
or Mott-insulating phase at $T=1/6$. It is seen that
the DOS around the Fermi level increases
as $V$ increases.  At $U=2$, the intermediate state
is first changed to the metallic state, where the quasi-particle 
peaks emerge in both orbitals ($V=0.75,1.0$). 
For fairly large $V$, the system falls into the 
renormalized band insulator ($V=1.5$).
In the case of $U=3$, the hybridization first drives the Mott-insulating
state to the intermediate one, as seen at $V=0.75$, which 
is followed by two successive transitions.

\begin{figure}[htb]
\begin{center}
\includegraphics[width=7cm]{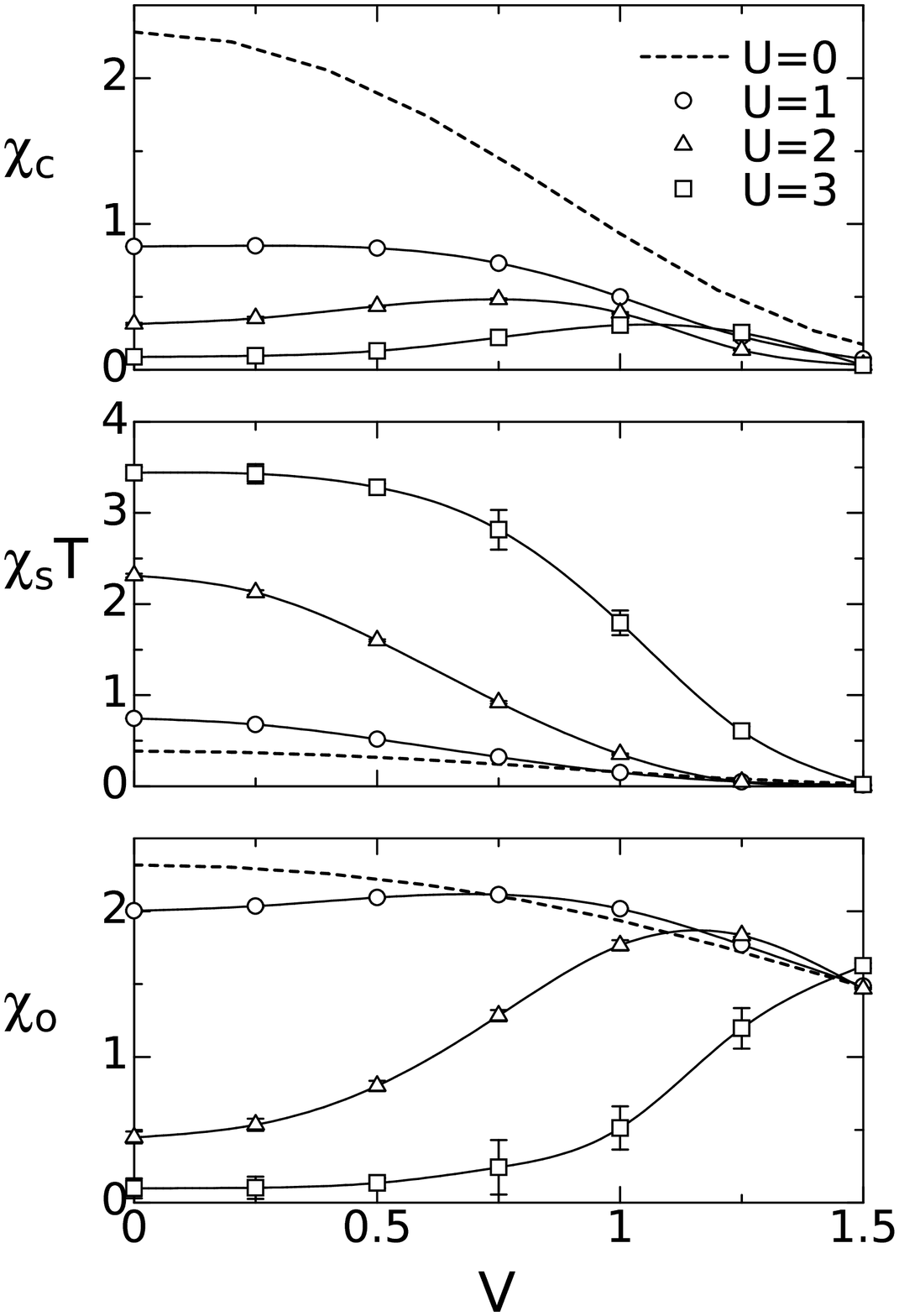}
\end{center}       
\caption{The charge, spin and orbital susceptibilities 
as a function of  $V$ at $T=1/6$.
}
\label{fig:chiall-v}
\end{figure}

These characteristic properties are also observed in the
charge, spin and orbital susceptibilities
at low temperatures (Fig. \ref{fig:chiall-v}).
For weak interactions ($U=1$), the charge 
susceptibility $\chi_c$ monotonically decreases with the increase of $V$.
When electron correlations become strong, 
nonmonotonic behavior appears in $\chi_c$:
the charge fluctuations, which are suppressed at $V=0$,
are somewhat recovered by the hybridization.
For large $V$, $\chi_c$ is suppressed again
since the system becomes a band insulator.
It is seen that the orbital susceptibility $\chi_o$ shows
nonmonotonic behavior similar to the charge susceptibility,
 the origin of which is essentially the
same as in $\chi_c$. In contrast, the spin susceptibility 
decreases with the increase of  $V$ irrespective of 
the strength of the interactions.  As is the case for
$V=0$, the effective spin is enhanced by ferromagnetic 
fluctuations due to the Hund coupling in the insulating 
and intermediate phases. When the hybridization is introduced  in 
these phases, the ferromagnetic fluctuations are 
suppressed, resulting in the monotonic decrease of the 
effective Curie constant.

\begin{figure}[htb]
\begin{center}
\includegraphics[width=6.3cm]{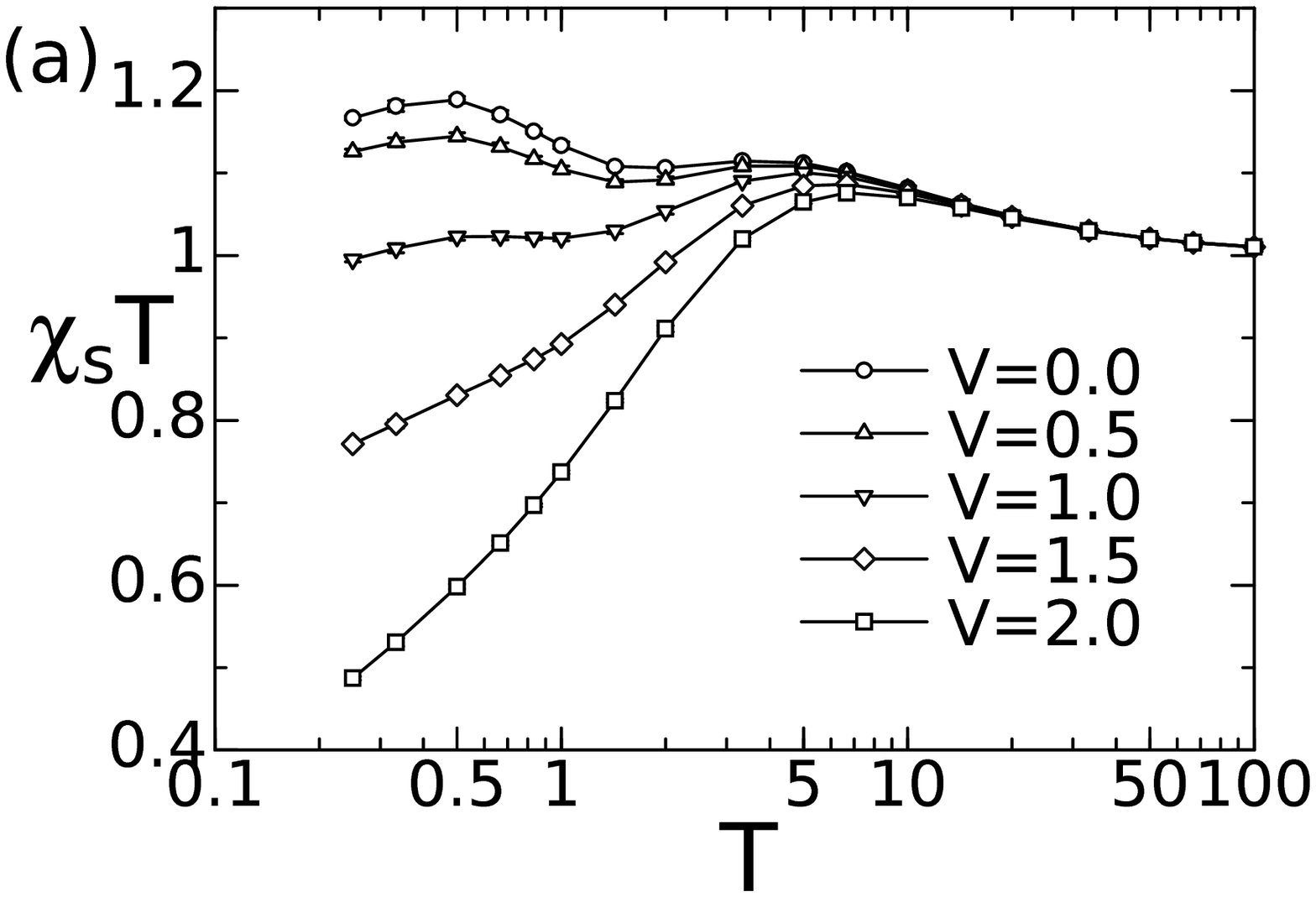}
\includegraphics[width=6cm]{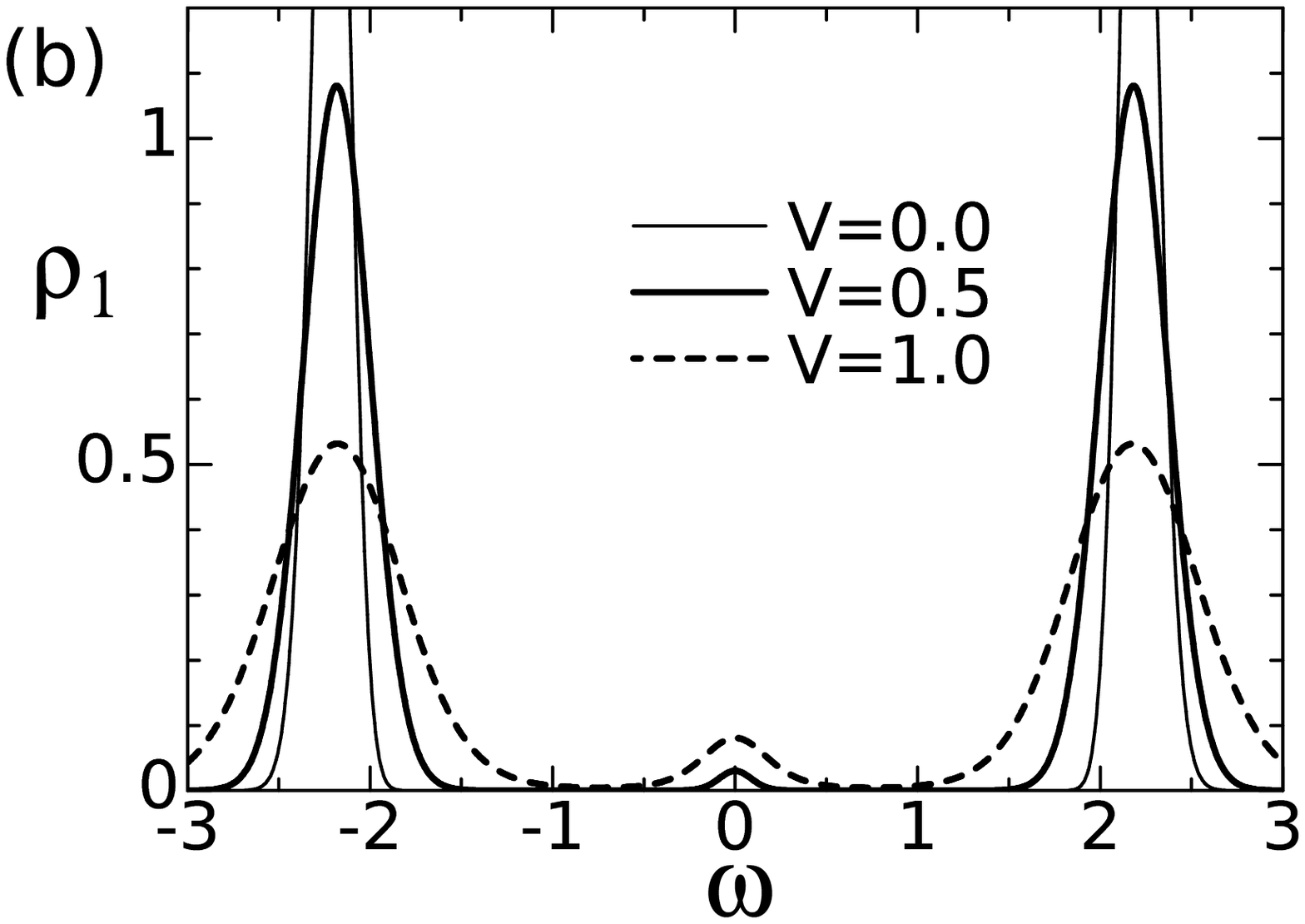}
\end{center}
\caption{(a) Effective Curie constant as a function of $T$
and (b) DOS in the narrower band $(\alpha=1)$  
at $T=1/4$  for
an extreme choice of the bandwidths, $(D_1,D_2)=(1.0, 10.0)$. 
The DOS for the wider band is not shown here.
The other parameters are $U=4.0, U'=3.0$ and $J=0.5$.
}
\label{fig:ex}
\end{figure}

We can thus say that the introduction of 
appropriate hybridization gives rise to  
heavy-fermion metallic behavior.
Such tendency is observed more clearly in an
extreme choice of the bandwidths, $(D_1, D_2)=(1.0, 10.0)$,
 as shown in Fig. \ref{fig:ex}. 
Since   the system is in the intermediate phase at $V=0.0$,
the narrower band shows localized-electron properties [Fig. \ref{fig:ex} (b)] 
in the background of 
the nearly free bands. This double structure in the DOS yields
two peaks in the temperature-dependent Curie constant,
as shown in Fig. \ref{fig:ex} (a). The localized state 
plays a role of the
$f$-state in the Anderson lattice model,
\cite{Kusunose} so that heavy-fermion quasi-particles appear
around the Fermi level for finite $V$, which are essentially the 
same as  those observed in Fig. \ref{fig:dos-V}.

\begin{figure}[htb]
\begin{center}
\includegraphics[width=7cm]{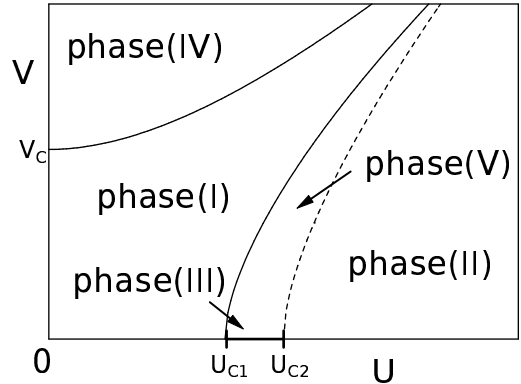}
\end{center}       
\caption{
The schematic phase diagram for the model 
with finite $V$ at $T=0$.
Solid lines represent the phase boundaries between the metallic and 
insulating phases. 
Dashed line indicates the phase boundary
between the Mott insulator and 
the Kondo insulator.
}
\label{fig:phase+}
\end{figure}

Finally, we make some comments on the phase diagram at $T=0$.
Although we have not yet studied low-temperature properties
in the presence of $V$, we can give some qualitative arguments on the 
 phase diagram expected at zero temperature.
As shown above, the metallic phase (I) is not so sensitive to 
$V$ in the weak-$V$ regime. This is also the case for the 
insulating phase (II),
where a triplet state ($S=1$) formed by the Hund coupling
is stable against a weak hybridization.
There appears a subtle situation 
in the intermediate phase (III). 
The intermediate phase exhibits Kondo-like heavy fermion
behavior at low temperatures in the presence of $V$. 
However,  we are now dealing with
 the half-filled case, so that this Kondo-like metallic phase 
should acquire a Kondo-insulating gap due to commensurability
at zero temperature. Therefore,  the intermediate
phase (III) should be changed into the Kondo-insulator with a tiny
excitation gap in the presence of $V$ at $T=0$. 
Accordingly, 
the sharp transition between the phases (II) and (III) at $V=0$
still remains in the weakly hybridized case. \cite{Medici05}
This is different from the situation in the periodic Anderson model
with the Coulomb interactions for conduction electrons 
as well as localized $f$ electrons,
where the spin-singlet insulating phase is always realized 
in its parameter range.\cite{Sato,SchorkPAM}
Consequently we end up with the schematic phase diagram 
(Fig. \ref{fig:phase+}) for 
the two-orbital model with the hybridization between the 
distinct orbitals.
Recall that the phase has three regions on the line $V=0$: 
(I) metallic, (II) insulating and
(III) intermediate phases.
The metallic phase (I) for small $U$ is simply driven to 
the band-insulator (IV) beyond  a certain critical value of $V$.
The intermediate phase (III) at $V=0$ is 
 changed to the Kondo-insulator (V)
in the presence of any finite $V$. As $V$ increases, this insulating state 
first undergoes a phase transition to the metallic phase (I), and
then enters the band-insulator (IV). On the other hand,
the Mott insulating phase (II) first shows a transition to
the Kondo insulator (V), which is further driven to the metallic phase (I)
and then to the band-insulating phase (IV).
Note that at finite temperatures above the Kondo-insulating gap,
we can observe Kondo-type heavy
fermion behavior in the intermediate phase with finite $V$.

\section{Summary}
We have investigated the Mott transitions in the two-orbital Hubbard model
by means of DMFT and SFA.
It has been clarified  that orbital fluctuations enhanced in the special
condition $U \sim U'$ and $J \sim 0$ play a  key role in 
stabilizing the correlated  metallic state.  In particular, this 
characteristic property gives rise to nontrivial effects on the 
Mott transition, which indeed control whether the OSMT is realized 
in the two-orbital model with different bandwidths.

We have demonstrated the above facts  by taking 
the system with the different bandwidths 
$(D_1, D_2)=(1.0, 2.0)$ as an example.  
In the special case with $U=U'$ and $J=0$, 
the metallic state is stabilized up to fairly large interactions,
resulting in a single Mott transition. The resulting single transition
is nontrivial, since we are concerned with the system
with different bandwidths. On the other hand,  for 
more general cases with $U\neq U'$ and $J\neq 0$,
the Hund coupling suppresses orbital fluctuations,
giving rise to the OSMT.  We have confirmed these results by 
computing  various quantities at zero and finite 
temperatures.

Recently, it was reported that 
when the ratio of the bandwidths is quite large,
the OSMT occurs even in the case of $U=U'$ and $J=0$.\cite{Medici05,Ferrero05}
This result implies that the detailed structure of the phase diagram in
some extreme cases depends on the parameters chosen, and is 
not completely categorized in the scheme discussed in this paper.
This naturally motivates us to establish the detailed phase diagrams 
at zero and finite temperatures by incorporating various effects such as
the magnetic field, the crystalline electric 
field,\cite{Ruegg}  the lattice structure, etc.

In this paper, we have restricted our discussions to the normal metallic 
and Mott insulating phases
without any long-range order. It is an important open problem to take into
account such instabilities to various ordered phases,
which is now under consideration.

\section*{Acknowledgement}
We are deeply indebted to our collaborators in this field, 
T. Ohashi, Y. Imai, S. Suga, T.M. Rice, and M. Sigrist, and
have benefitted from helpful discussions with
F. Becca, S. Biermann, A. Georges, A. Liebsch, S. Nakatsuji, 
and Y. Maeno. 
Discussions during the YITP workshop YKIS2004 on 
"Physics of Strongly Correlated Electron Systems" 
were useful to complete this work.
This work was partly supported by a Grant-in-Aid from the Ministry 
of Education, Science, Sports and Culture of Japan, 
the Swiss National Foundation and the NCCR MaNEP.
A part of computations was done at the Supercomputer Center at the 
Institute for Solid State Physics, University of Tokyo
and Yukawa Institute Computer Facility. 



\begin{thebibliography}{999}




\bibitem{ImadaRev}
M. Imada, A. Fujimori and Y. Tokura, Rev. Mod. Phys. {\bf 70}, 1039 (1998).

\bibitem{TokuraScience}
Y. Tokura and N. Nagaosa, Science {\bf 288},462 (2000).


\bibitem{Georges}
A. Georges, G. Kotliar, W. Krauth and M. J. Rozenberg,
Rev. Mod. Phys. {\bf 68}, 13 (1996).

\bibitem{Metzner}
W. Metzner and D. Vollhardt, Phys. Rev. Lett. {\bf 64}, 324 (1989).

\bibitem{Muller}
E. M\"uller-Hartmann, Z. Phys. B, Condens. Matter {\bf 74}, 507 (1989).

\bibitem{PruschkeAP}
T. Pruschke, M. Jarrell and J. K. Freericks, Adv. Phys. 44, 187 (1995).

\bibitem{KotliarPT}
G. Kotliar and D. Vollhardt, Physics Today 53, (2004).




\bibitem{Liebsch}
A. Liebsch, Europhys. Lett., {\bf 63}, 97 (2003); 
Phys. Rev. Lett., {\bf 91}, 226401 (2003).

\bibitem{sces}
A. Koga, N. Kawakami, T. M. Rice, and M. Sigrist, 
Physica B {\bf 359-361}, 1366 (2005).

\bibitem{Knecht05}
C. Knecht, N. Bl{\"u}mer and P. G. J. van Dongen, 
Phys. Rev. B {\bf 72}, 081103 (2005).

\bibitem{LiebschFiniteT}
A. Liebsch, Phys. Rev. Lett. {\bf 95}, 116402 (2005).

\bibitem{Biermann05}
S. Biermann, L. de' Medici and A. Georges, cond-mat/0505737.

\bibitem{Ruegg}
A. R\"uegg, M. Indergand, S. Pilgram, and M. Sigrist, cond-mat/0508691.


\bibitem{KogaLett}
A. Koga, N. Kawakami, T.M. Rice and M. Sigrist, 
Phys. Rev. Lett. {\bf 92}, 216402 (2004)

\bibitem{KogaB}
A. Koga, N. Kawakami, T. M. Rice, and M. Sigrist, 
Phys. Rev. B {\bf 72}, 045128 (2005).

\bibitem{Tomio04}
Y. Tomio and T. Ogawa, 
J. Luminesci. {\bf 112}, 220 (2005).

\bibitem{Ferrero05}
M. Ferrero, F. Becca, M. Fabrizio and M. Capone, cond-mat/0503759.

\bibitem{Medici05}
L. de' Medici, A. Georges and S. Biermann, cond-mat/0503764.

\bibitem{Arita05}
R. Arita and K. Held, cond-mat/0504040.

\bibitem{Inaba}
K. Inaba, A. Koga, S. Suga, N. Kawakami, 
J. Phys. Soc. Jpn. {\bf 74}, 2393 (2005).


\bibitem{2band1}
A. Georges, G. Kotliar and W. Krauth, Z. Phys. B {\bf 92}, 313 (1993).

\bibitem{Kotliar96}
G. Kotliar and H. Kajueter, Phys. Rev. B {\bf 54}, 14221 (1996).

\bibitem{Rozenberg97}
M. J. Rozenberg, Phys. Rev. B {\bf 55}, 4855 (1997).

\bibitem{Bunemann:Gutzwiller}
J. B\"unemann and W. Weber, Phys. Rev. B {\bf 55}, 4011 (1997); 
J. B\"unemann and W. Weber and F. Gebhard, 
Phys. Rev. B {\bf 57}, 6896 (1998).

\bibitem{Hasegawa98}
H. Hasegawa, J. Phys. Soc. Jpn. {\bf 56}, 1196 (1997).

\bibitem{Held98}
K. Held and D. Vollhardt, Eur. Phys. J. B {\bf 5}, 473 (1998).

\bibitem{Han98}
J. E. Han, M. Jarrel and D. L. Cox, Phys. Rev. B {\bf 58}, 4199 (1998).

\bibitem{Momoi98}
T. Momoi and K. Kubo, Phys. Rev. B {\bf 58}, 567 (1998).

\bibitem{Klejnberg98}
A. Klejnberg and J. Spalek, Phys. Rev. B {\bf 57}, 12401 (1998).

\bibitem{2band2}
Th. Maier, M. B. Z\"olfl, Th. Pruschke and J. Keller, 
Eur. Phys. J. B {\bf 7}, 377 (1999).

\bibitem{Imai01}
Y. Imai and N. Kawakami, J. Phys. Soc. Jpn {\bf 70}, 2365 (2001).

\bibitem{Oudovenko02}
V. S. Oudovenko and G. Kotliar, Phys. Rev. B {\bf 65}, 075102 (2002).

\bibitem{Koga}
A. Koga, Y. Imai and N. Kawakami, Phys. Rev. B {\bf 66}, 165107 (2002).

\bibitem{Florens}
S. Florens, A. Georges,  G. Kotliar and O. Parcollet,
Phys. Rev. B {\bf 66}, 205102 (2002);
S. Florens, A. Georges, Phys. Rev. B {\bf 66}, 165111 (2002).

\bibitem{Ono03}
Y. Ono, M. Potthoff and R. Bulla, Phys. Rev. B {\bf 67}, 035119 (2003).

\bibitem{KogaSN}
A. Koga, T. Ohashi, Y. Imai, S. Suga and N. Kawakami, 
J. Phys. Soc. Jpn. {\bf 72}, 1306 (2003).

\bibitem{Sakai}
S. Sakai, R. Arita, and H. Aoki, Phys. Rev. B {\bf 70}, 172504 (2004).

\bibitem{Pruschke}
T. Pruschke and R. Bulla, Eur. Phys. J. B {\bf 44}, 217 (2005).

\bibitem{Song}
Y. Song and L.-J. Zou, 
Phys. Rev. B {\bf 72}, 085114 (2005).

\bibitem{InabaB}
K. Inaba, A. Koga, S. Suga, and N. Kawakami, 
Phys. Rev. B {\bf 72}, 085112 (2005).



\bibitem{Anisimov}
V.I. Anisimov, I.A. Nekrasov, D.E. Kondakov, T.M. Rice and M. Sigrist,
Eur. Phys. J. B {\bf 25}, 191 (2002).

\bibitem{Nakatsuji1}
S. Nakatsuji and Y. Maeno, Phys. Rev. Lett. {\bf 84}, 2666 (2000).

\bibitem{tilting}
O. Friedt, M. Braden, G. Andr\'e, P. Adelmann, S. Nakatsuji, and Y. Maeno,
Phys. Rev. B {\bf 63}, 174432 (2001). 

\bibitem{Nakatsuji}
S. Nakatsuji, D. Hall, L. Balicas, Z. Fisk, K. Sugahara, M. Yoshioka,
and Y. Maeno,
Phys. Rev. Lett. {\bf 90}, 137202 (2003).

\bibitem{PT} 
Y. Maeno, T.M. Rice and M. Sigrist, Physics Today, 
{\bf 54}, 42 (2001).

\bibitem{RMP}
A.P. Mackenzie and Y. Maeno, Rev. Mod. Phys. {\bf 75}, 657 (2003).

\bibitem{Ca2RuO41}
S. Nakatsuji, S. I. Ikeda, and Y. Maeno, 
J. Phys. Soc. Jpn. {\bf 66}, 1868 (1997).

\bibitem{Ca2RuO42}
M. Braden, G. Andr\'e, S. Nakatsuji and Y. Maeno,
Phys. Rev. B {\bf 58}, 847 (1998).

\bibitem{Ca2RuO43}
F. Nakamura, T. Goko, M. Ito, T. Fujita, S. Nakatsuji, 
H. Fukazawa, Y. Maeno, P. Alireza, D. Forsythe, and S. R. Julian,
Phys. Rev. B {\bf 65}, 220402 (2002).

\bibitem{Mazin}
I. I. Marzin and D. J. Singh, Phys. Rev. Lett. {\bf 82}, 4324 (1999).

\bibitem{Hotta}
T. Hotta and E. Dagotto, Phys. Rev. Lett. {\bf 88}, 017201 (2001).

\bibitem{Fang}
Z. Fang and K. Terakura, Phys. Rev. B {\bf 64}, R020509 (2001);
Z. Fang, N. Nagaosa and K. Terakura, 
Phys. Rev. B {\bf 69}, 045116 (2004).

\bibitem{Okamoto}
S. Okamoto and A. J. Mills, Phys. Rev. B {\bf 70}, 195120 (2004).

\bibitem{SigristTroyer}
M. Sigrist and M. Troyer, Eur. J. Phys. B, {\bf 39}, 207 (2004).

\bibitem{LaNiO}
K. Sreedhar, M. McElfresh, D. Perry, D. Kim, P. Metcalf and J. M. Honig,
J. Solid State Comm. {\bf 110}, 208 (1994); 
Z. Zhang, M. Greenblatt and J. B. Goodenough,
J. Solid State Comm. {\bf 108}, 402 (1994); {\bf 117}, 236 (1995).

\bibitem{Kobayashi96}
Y. Kobayashi, S. Taniguchi, M. Kasai, M. Sato, T. Nishioka and M. Kontani,
J. Phys. Soc. Jpn {\bf 65}, 3978 (1996).



\bibitem{SFA}
M. Potthoff, Eur. Phys. J. B {\bf 32}, 429 (2003); {\bf 36}, 335 (2003); 
K. Pozg{\v{a}j\v{c}i\'{c}}, cond-mat/0407172.




\bibitem{Kuramoto}
Y. Kuramoto, Z. Phys. B {\bf 53}, 37 (1983).

\bibitem{Coleman}
P. Coleman, Phys. Rev. B {\bf 28}, 5255 (1983); {\bf 29}, 3035 (1984).

\bibitem{Rice}
T.M. Rice and K. Ueda, Phys. Rev. Lett. {\bf 55}, 995 (1985);
Phys. Rev. B {\bf 34}, 6420 (1986).

\bibitem{Kim90}
C.-I. Kim, Y. Kuramoto and T. Kasuya, J. Phys. Soc. Jpn. {\bf 59}, 2414 (1990).

\bibitem{Yamada}
K. Yamada, K. Yosida, and K. Hanzawa, 
Prog. Thero, Phys. Suppl. {\bf 108}, 141 (1992).


\bibitem{Tsunetsugu}
H. Tsunetsugu, M. Sigrist and K. Ueda, 
Rev. Mod. Phys. {\bf 69}, 809 (1997).

\bibitem{Assaad}
F. F. Assaad, Phys. Rev. Lett. {\bf 83}, 796 (1999).


\bibitem{Zener}
C. Zener, Phys. Rev. {\bf 82}, 4031 (1951).

\bibitem{Anderson}
P. W. Anderson and H. Hasegawa, Phys. Rev. {\bf 100}, 675 (1955).

\bibitem{Kubo}
K. Kubo and N. Ohata, J. Phys. Soc. Jpn. {\bf 33}, 21 (1975).

\bibitem{Furukawa}
N. Furukawa, J. Phys. Soc. Jpn. {\bf 64}, 2734 (1995).



\bibitem{single1}
Th. Pruschke, D. L. Cox and M. Jarrell, Phys. Rev. B {\bf 47}, 3553 (1993).

\bibitem{OSakai}
O. Sakai and Y. Kuramoto, Solid State Comm. {\bf 89}, 307 (1994).

\bibitem{Caffarel}
M. Caffarel and W. Krauth, Phys. Rev. Lett. {\bf 72}, 1545 (1994).  

\bibitem{Rozenberg1}
M. J. Rozenberg, G. Kotliar, and X. Y. Zhang,
Phys. Rev. B {\bf 49}, 10181 (1994).

\bibitem{BullaNRG}
R. Bulla, Phys. Rev. Lett. {\bf 83}, 136 (1999);
R. Bulla, T. A. Costi, D. Vollhardt, Phys. Rev. Lett. {\bf 64}, 045103 (2001).

\bibitem{single2}
R. Chitra and G. Kotliar, Phys. Rev. Lett. {\bf 83}, 2386 (1999).

\bibitem{LDMFT}
R. Bulla and M. Potthoff, Eur. Phys. J. B {\bf 13}, 257 (2000).

\bibitem{2site}
M. Potthoff, Phys. Rev. B {\bf 64}, 165114 (2001).

\bibitem{OnoED}
Y. Ono, R. Bulla and A. C. Hewson, Eur. Phys. J. B {\bf 19}, 375 (2001);
Y. Ohashi and Y. Ono, J. Phys. Soc. Jpn. {\bf 70}, 2989 (2001).

\bibitem{single3}
J. Joo and V. Oudovenko, Phys. Rev. B {\bf 64}, 193102 (2001).

\bibitem{single4}
M. S. Laad, L. Craco and E. M\"uller-Hartmann,
Phys. Rev. B {\bf 64}, 195114 (2001). 

\bibitem{Nishimoto}
S. Nishimoto, F. Gebhard, and E. Jeckelmann,
J. Phys. Condens. Matter {\bf 16}, 7063 (2004).

\bibitem{Zitzler}
R. Zitzler, N.-H. Tong, Th. Pruschke, and R. Bulla,
Phys. Rev. Lett. {\bf 93}, 016406 (2004).

\bibitem{Uhrig}
M. Karski, C. Raas, and G. S. Uhrig,
cond-mat/0507132.


\bibitem{PAM}
M. Jarrell, H. Akhlaghpour and T. Pruschke,
Phys. Rev. Lett. {\bf 70}, 1670 (1993). 

\bibitem{Mutou}
T. Mutou and D. Hirashima, J. Phys. Soc. Jpn. {\bf 63}, 4475 (1994);

\bibitem{Rozenberg}
M. J. Rozenberg, Phys. Rev. B {\bf 52}, 7369 (1995).

\bibitem{Saso}
T. Saso and M. Itoh, Phys. Rev. B {\bf 53}, 6877 (1996);  

\bibitem{SchorkPAM}
T. Schork and S. Blawid, Phys. Rev B {\bf 56}, 6559 (1997)

\bibitem{Sun}
P. Sun and G. Kotliar, Phys. Rev. B {\bf 91}, 037209 (2003).

\bibitem{Ohashi}
T. Ohashi, A. Koga, S. Suga, and N. Kawakami, 
Phys. Rev. B {\bf 70}, 245104 (2004);
Physica B {\bf 359-361}, 738 (2005).

\bibitem{Sato}
R. Sato, T. Ohashi, A. Koga, and N. Kawakami, 
J. Phys. Soc. Jpn. {\bf 73}, 1864 (2004).

\bibitem{MediciPAM}
L. de' Medici, A. Georges, G. Kotliar, and S. Biermann, 
Phys. Rev. Lett. {\bf 95}, 066402 (2005).


\bibitem{Matsumoto}
N. Matsumoto and F. J. Ohkawa,
Phys. Rev. B {\bf 51}, 4110 (1995).

\bibitem{Schork}
T. Schork, S. Blawid, and J. Igarashi,
Phys. Rev. B {\bf 59}, 9888 (1999).

\bibitem{OhashiJPC}
T. Ohashi, S. Suga, and N. Kawakami,
J. Phys. Condens. Matter {\bf 17}, 4547 (2005).

\bibitem{Hirsch}
J. E. Hirsch and R. M. Fye, Phys. Rev. Lett. {\bf 56}, 2521 (1986).

\bibitem{Luttinger}
J. M. Luttinger, Phys. Rev. {\bf 118}, 1417 (1960).

\bibitem{MEM1}
S.F. Gull, in Maximum Entropy and Bayesian Methods in Science and Engineering, 
edited by G. J. Erickson and C. R. Smith 
(Kluwer Academic, Dordrecht, 1988), p. 53; J. Skilling, p. 45.

\bibitem{MEM2}
R. N. Silver, D. S. Sivia, and J. E. Gubernatis, 
Phys. Rev. B {\bf 41}, 2380 (1990); 
J. E. Gubernatis, M. Jarrell, R. N. Silver, and D. S. Sivia, 
Phys. Rev. B {\bf 44}, 6011 (1991).

\bibitem{MEM3}
W.F. Press, S.A. Teukolsky, W.T. Vetterling, and B.R. Flannery, 
Numerical Recipes 
(Cambridge University Press, Cambridge, England, 1992), p. 809.

\bibitem{Gutzwiller}
M. C. Gutzwiller, Phys. Rev. {\bf 137}, A1726 (1965).

\bibitem{Brinkman}
W. F. Brinkman and T. M. Rice, Phys. Rev. B {\bf 2}, 4302 (1970).

\bibitem{Kusunose}
H. Kusunose, S. Yotsuhashi, and K. Miyake, Phys. Rev. B {\bf 62}, 4403 (2000).




\end{thebibliography}
\end{document}